%% file: bare_jrnl.tex
\documentclass[journal]{IEEEtran}
\usepackage[numbers]{natbib}

\usepackage{booktabs}
\usepackage{tabularx}
\usepackage{array}
\usepackage{multirow}
\usepackage{makecell}
\usepackage{longtable}
\usepackage{xltabular}
\usepackage{changepage}
\usepackage{amssymb}
\usepackage{xcolor}
\usepackage{tikz}
\usepackage[most]{tcolorbox}
\usepackage{caption}
\usepackage{float}
\usepackage{placeins}
\usepackage{xurl}
\usepackage{hyperref}
\usepackage{colortbl}

\newcommand{\concept}[1]{%
  \par\noindent\textbf{#1}\enspace
}

\newcommand{\archbadge}[2]{%
    \tikz[baseline=(char.base)]{
        \node[
            circle,
            fill=#1,
            text=white,
            minimum size=4.3mm,
            inner sep=0pt,
            font=\bfseries\tiny
        ] (char) {#2};
    }%
}

\newcommand{\archS}{\archbadge{black!90}{S}}
\newcommand{\archH}{\archbadge{blue!75!black}{H}}
\newcommand{\archP}{\archbadge{green!60!black}{P}}
\newcommand{\archC}{\archbadge{orange!85!black}{C}}
\newcommand{\archD}{\archbadge{violet!75!black}{D}}
\newcommand{\archM}{\archbadge{cyan!65!black}{M}}

\newcommand{\archsep}{\hspace{1.5pt}}

\newcommand{\artbadge}[2]{%
    \tikz[baseline=(char.base)]{
        \node[
            rounded corners=1pt,
            fill=#1,
            text=white,
            minimum width=4.3mm,
            minimum height=4.1mm,
            inner sep=0pt,
            font=\bfseries\tiny
        ] (char) {#2};
    }%
}

\newcommand{\artC}{\artbadge{green!40!blue!75!black}{C}}
\newcommand{\artP}{\artbadge{magenta!75!black}{P}}
\newcommand{\artD}{\artbadge{yellow!55!black}{D}}
\newcommand{\artE}{\artbadge{red!75!black}{E}}

\newcommand{\artsep}{\hspace{1.2pt}}

\definecolor{coverageblue}{RGB}{48,92,135}
\definecolor{coveragelight}{RGB}{242,246,249}
\definecolor{coveragegray}{RGB}{205,210,214}

\newcommand{\covered}{%
    \tikz[baseline=(char.base)]{
        \node[
            circle,
            fill=coverageblue,
            text=white,
            minimum size=4.5mm,
            inner sep=0pt,
            font=\bfseries\tiny
        ] (char) {\checkmark};
    }%
}

\newcommand{\notcovered}{%
    \tikz[baseline=(char.base)]{
        \node[
            circle,
            draw=coveragegray,
            fill=white,
            minimum size=4.5mm,
            inner sep=0pt
        ] (char) {};
    }%
}

\newcommand{\atkB}{\textsf{\textbf{B}}}
\newcommand{\atkG}{\textsf{\textbf{G}}}
\newcommand{\atkW}{\textsf{\textbf{W}}}

\newcommand{\defyes}{\textcolor{green!45!black}{\checkmark}}
\newcommand{\defno}{--}

\newcommand{\repolink}[1]{%
    \href{#1}{\textsuperscript{\textsf{[R]}}}%
}

\begin{document}
\let\WriteBookmarks\relax
\def\floatpagepagefraction{1}
\def\textpagefraction{.001}

\def\tsc#1{\csdef{#1}{\textsc{\lowercase{#1}}\xspace}}
\tsc{WGM}
\tsc{QE}
\tsc{EP}
\tsc{PMS}
\tsc{BEC}
\tsc{DE}

\newcolumntype{Y}{>{\raggedright\arraybackslash}X}

\newtcolorbox{openquestion}[1]{
    colback=blue!12,
    colframe=blue!70!black,
    boxrule=0.4pt,
    arc=1pt,
    left=4pt,
    right=4pt,
    top=3pt,
    bottom=3pt,
    before skip=4pt,
    after skip=4pt,
    fonttitle=\bfseries,
    title=#1
}
\newtcolorbox{rqanswer}[1]{
  colback=gray!6,
  colframe=black!45,
  boxrule=0.5pt,
  arc=1pt,
  left=5pt,
  right=5pt,
  top=4pt,
  bottom=4pt,
  title=\textbf{#1},
  fonttitle=\small,
  coltitle=black,
  colbacktitle=gray!15,
  before skip=6pt,
  after skip=8pt
}

\ifCLASSINFOpdf
\else
\fi
\title{Connecting the Dots in Agentic AI Security: A Cross-Dimensional Threat Taxonomy, Evaluation Maturity, and Open Challenges}

%
%
%

\author{Heewon Baek,
        Alsharif Abuadbba,
        Kristen Moore, 
        Hyoungshick Kim,
        Surya Nepal
\thanks{Heewon Baek and Hyoungshick Kim are with Sungkyunkwan University, Republic of Korea. The work was conducted during Heewon's internship at CSIRO, Australia}
\thanks{Alsharif Abuadbba,  Kristen Moore and Surya Nepal are with Commonwealth Scientific and Industrial Research Organisation (CSIRO), Australia.}
}

\markboth{}{}

\maketitle

\begin{abstract}
Agentic AI extends LLM security beyond generated content to persistent
state, autonomous actions, tool use, and interactions with humans and
other agents. Existing threat classifications often emphasize individual
dimensions, obscuring connections among entry points, affected components,
and security consequences. The known threat landscape also differs from
the coverage demonstrated by empirical research.

Through a structured review of 66 studies published from 2022 to 2026,
we introduce $\mathcal{T}=\langle \mathcal{S},\mathcal{B},\mathcal{P},\mathcal{A}\rangle$, a cross-dimensional
representation linking affected functional or system surfaces ($\mathcal{S}$),
interaction or trust boundaries ($\mathcal{B}$), violated security properties ($\mathcal{P}$),
and empirically examined architectures ($\mathcal{A}$).
We analyze 22 artifact-backed red-teaming studies and 11 representative
security benchmarks to characterize empirical coverage and evaluation
maturity. Within the selected studies, evidence concentrates on
prompt/reasoning, memory, and tool-mediated attacks, predominantly in
single-agent settings. Persistent, Human--Agent, complex multi-agent,
systemic, and long-horizon threats receive less coverage.
These findings describe the selected corpus rather than establish gaps
across all empirical research. Heterogeneous metrics, limited adaptive
defense evaluation, architectural imbalance, and incomplete execution-state
capture further constrain comparison and reproducibility.
We derive 13 open research questions to guide more systematic,
architecture-aware, and reproducible security evaluation of agentic AI.
\end{abstract}


%
\IEEEpeerreviewmaketitle

\input{Section/introduction}
\input{Section/review_methodology}
\input{Section/background}
\input{Section/taxonomy_of_security_threats}
\input{Section/red-teaming}
\input{Section/discussion}
\input{Section/conclusion}

\ifCLASSOPTIONcaptionsoff
  \newpage
\fi

\bibliographystyle{IEEEtran}

\bibliography{References}

\end{document}

%% file: Section/introduction.tex
\section{Introduction}
\label{sec:introduction}

Large language models (LLMs) are increasingly being embedded into
autonomous systems that can reason over goals, construct and revise plans,
maintain persistent state, invoke external tools, and interact with users,
environments, and other agents. These systems, commonly referred to as
\emph{agentic AI}, extend the capabilities of conventional LLMs from
generating responses to selecting and executing actions over time
\cite{wang2024survey,xi2025rise}. Contemporary agent architectures combine
LLMs with components such as planning and reflection mechanisms, short- and
long-term memory, retrieval systems, tool interfaces, and multi-agent
coordination \cite{wu2024autogen,yao2022react,shinn2023reflexion}. As a
result, an agent is no longer merely a model responding to an input, but a
stateful decision-making system whose outputs can alter digital and, in
increasingly important settings, physical environments.

This transition fundamentally changes the security problem. In a
conventional LLM interaction, an adversary primarily attempts to manipulate
the information processed or generated by the model. In an agentic system,
the consequences of such manipulation can propagate through an
\emph{action loop}: an adversarial input can alter reasoning, the resulting
decision can modify memory or invoke a tool, the tool can change an external
state, and that state can subsequently be observed by the agent and influence
future decisions. Moreover, external information retrieved from websites,
documents, memories, tools, or other agents can itself carry adversarial
instructions or poisoned state. Security therefore depends not only on the
robustness of the underlying LLM, but also on the integrity of goals and
plans, persistent state, authorization boundaries, tool and software supply
chains, inter-agent trust, human oversight, and the environments in which
agents operate. Recent attacks involving indirect prompt injection, memory
poisoning, malicious tools, compromised tool responses, privilege misuse,
and adversarial multi-agent interactions illustrate different manifestations
of this expanded attack surface
\cite{greshake2023not,debenedetti2024agentdojo,chen2024agentpoison,
he2025red,abuadbba2026human,singh2026shifting}.

\begin{table*}[!t]
\caption{Positioning of this survey relative to representative academic and
industry-oriented work on agentic AI security. Filled markers indicate
substantive coverage of the corresponding dimension.}
\label{tab:related_surveys}

\centering
\small
\renewcommand{\arraystretch}{1.30}
\setlength{\tabcolsep}{4.5pt}

\resizebox{0.98\textwidth}{!}{%
\begin{tabular}{
    >{\raggedright\arraybackslash}p{4.0cm}
    *{7}{>{\centering\arraybackslash}p{2.15cm}}
}
\toprule

&
\multicolumn{4}{c}{\textbf{Threat and System Characterization}} &
\multicolumn{3}{c}{\textbf{Empirical Evaluation}} \\

\cmidrule(lr){2-5}
\cmidrule(lr){6-8}

\textbf{Work} &
\makecell[c]{\textbf{Threat}\\\textbf{taxonomy}} &
\textbf{Defenses} &
\makecell[c]{\textbf{Agent}\\\textbf{architecture}} &
\makecell[c]{\textbf{Trust/interaction}\\\textbf{boundaries}} &
\makecell[c]{\textbf{Red-team evidence}\\\textbf{mapping}} &
\makecell[c]{\textbf{Benchmarks /}\\\textbf{evaluation}} &
\makecell[c]{\textbf{Reproducibility}\\\textbf{analysis}} \\

\midrule

Deng et al.~\cite{deng2025ai}
&
\covered & \covered & \covered & \covered &
\notcovered & \notcovered & \notcovered
\\

Shahriar et al.~\cite{shahriar2025agentic}
&
\covered & \covered & \covered & \notcovered &
\notcovered & \notcovered & \notcovered
\\

Chhabra et al.~\cite{chhabra2026agentic}
&
\covered & \covered & \covered & \covered &
\notcovered & \covered & \notcovered
\\

Kim et al.~\cite{kim2026attack}
&
\covered & \covered & \covered & \covered &
\notcovered & \covered & \notcovered
\\

OWASP Agentic Security~\cite{owasp_agentic_2025}
&
\covered & \covered & \notcovered & \covered &
\notcovered & \notcovered & \notcovered
\\

MITRE ATLAS~\cite{mitre_atlas}
&
\covered & \covered & \notcovered & \covered &
\notcovered & \notcovered & \notcovered
\\

\midrule

\rowcolor{coveragelight}
\textbf{This survey}
&
\covered & \covered & \covered & \covered &
\covered & \covered & \covered
\\

\bottomrule
\end{tabular}%
}



\end{table*}

The rapid emergence of these threats has produced a growing body of research
that attempts to organize agent security. Deng et al.
\cite{deng2025ai} characterize the security challenges of AI agents through
four key gaps concerning multi-step user inputs, internal execution,
operational environments, and interactions with untrusted entities.
Subsequent surveys broaden this perspective. Shahriar et al.
\cite{shahriar2025agentic} organize the literature around applications,
threats, and defenses, while Chhabra et al.
\cite{chhabra2026agentic} jointly examine threats, defenses, evaluation
methods, and open challenges. More recently, Kim et al.
\cite{kim2026attack} analyze the agent design space together with its attack
and defense landscape. In parallel, practitioner-oriented frameworks such as
the OWASP Agentic Security Initiative and MITRE ATLAS provide operational
threat knowledge intended to support threat modeling and security assessment
\cite{owasp_agentic_2025,mitre_atlas}. Collectively, these efforts have
substantially improved our understanding of \emph{what threats may exist} in
agentic AI systems. Table~\ref{tab:related_surveys} summarizes the positioning of this survey
relative to representative academic surveys and practitioner-oriented
frameworks, highlighting the distinction between threat and system
characterization and empirical security evaluation.

However, two complementary gaps remain. First, existing classifications
provide valuable views of agentic AI security but commonly organize threats
primarily by individual dimensions, such as agent component, attack
mechanism, lifecycle stage, security objective, or operational risk
~\cite{wang2026understanding}. These dimensions are not independent in
agentic systems. An attack may enter through one trust boundary, compromise
a different functional component, persist or propagate through the system state,
and ultimately violate a security property elsewhere in the execution
trajectory. For example, adversarial content retrieved from an external
tool may compromise reasoning, persist in memory, and later induce an
unauthorized action. \textit{Characterizing such threats along a single dimension
can therefore obscure the path of compromise and the role of agent
architecture in shaping its propagation and impact.}

Second, the breadth of the \emph{known threat landscape} should not be
conflated with the breadth of the \emph{empirically evaluated threat
landscape}. Threats may be identified in surveys, threat models, or
practitioner frameworks without being systematically instantiated against
operational agents. Conversely, empirical studies vary substantially in
attacker assumptions, target architectures, execution environments,
benchmarks, metrics, defense evaluation, and reproducibility artifacts.
Understanding agentic AI security therefore requires not only identifying
\emph{what can go wrong}, but also determining \emph{what has actually been
tested} and \emph{how mature the supporting evidence is}.

Motivated by these gaps, this survey \emph{connects the dots} between the
structure of the agentic AI attack surface and the empirical evidence
available to support it. Rather than viewing threats through \textit{a single
dimension}---such as attack type, system component, trust boundary, or
security objective---we adopt a \emph{cross-dimensional} view that captures
how compromise traverses an agentic system. Specifically, we cross-map each
threat by the functional or system surface affected, the interaction or trust
boundary through which compromise enters or propagates, the security property
violated, and the architecture in which the threat has been empirically
examined. This mapping connects \emph{where compromise manifests}, \emph{how
it propagates}, and \emph{what consequence it produces} with evidence of
\emph{where it has actually been tested}. We then map empirical red-teaming
studies and security benchmarks onto this representation to determine which
regions of the broader threat landscape have been evaluated and how mature
the supporting evidence is. Specifically, we address the following research
questions:

\begin{itemize}
    \item[\textbf{RQ1}] \textbf{Attack Surface:} How can security threats in
    agentic AI systems be systematically characterized across functional
    components, architectural interactions, and trust boundaries?

    \item[\textbf{RQ2}] \textbf{Empirical Coverage:} Which regions of this
    attack surface have been empirically investigated by agentic AI
    red-teaming research, and under what adversarial assumptions and
    experimental settings?

    \item[\textbf{RQ3}] \textbf{Evaluation Maturity:} How mature is the
    empirical evidence supporting current agent-security claims in terms of
    benchmarks, metrics, realistic execution environments, defenses, and
    reproducibility?
\end{itemize}

To answer these questions, we make the following contributions:

\begin{itemize}

\item \textbf{A cross-dimensional security framework.}
Drawing on 66 agentic AI security studies from 2022--2026, we connect
affected functional or system surfaces, interaction or trust boundaries,
violated security properties, and empirically examined architectures.
This representation links complementary dimensions of threats across
components and system boundaries.

\item \textbf{An evidence map of agentic AI red teaming.}
We map demonstrated attacks from 22 artifact-backed empirical studies
onto the framework, relating the known threat landscape to the coverage
observed within this subset across threats, attacker assumptions,
architectures, and experimental settings.

\item \textbf{An assessment of evaluation maturity.}
We examine these studies alongside 11 representative security benchmarks
and evaluation environments, assessing coverage, metrics, execution
realism, defense evaluation, and public artifacts. We identify limitations
in architectural diversity, temporal scope, outcome comparability, and
reproducibility support within the reviewed evidence.

\item \textbf{An evidence-driven research agenda.}
We derive 13 open research questions from the observed limitations,
covering runtime and persistent security, multi-agent and Human--Agent
interactions, supply-chain and cyber--physical risks, and reproducible
system-level evaluation.

\end{itemize}

The survey highlights the need to complement \emph{threat enumeration}
with \emph{systematic security evaluation}. By linking threat structure
to empirical coverage and evaluation maturity, it clarifies what the
reviewed evidence supports and where further investigation is warranted.
These findings characterize the selected studies rather than establish
the absence of evidence in the broader literature.

\concept{Roadmap.}
Section~\ref{sec:methodology} describes the literature-review and
evidence-selection methodology. Section~\ref{sec:background} establishes the agentic AI system and
architectural model used throughout the survey.
Section~\ref{sec:taxonomy} develops the cross-dimensional threat framework
and characterizes the broader attack surface (RQ1).
Section~\ref{sec:redteaming} maps empirical red-teaming evidence and
security benchmarks onto this framework to assess empirical coverage (RQ2)
and evaluation maturity (RQ3).
Section~\ref{sec:discussion} derives the resulting research gaps and open
questions, and Section~\ref{sec:conclusion} concludes the survey.

%% file: Section/review_methodology.tex
\section{Review Methodology}
\label{sec:methodology}

To provide a transparent and reproducible basis for the survey, we adopt a
structured literature-review methodology aligned with the three research
questions introduced in Section~\ref{sec:introduction}. The review consists
of two related stages. First, we construct a broad corpus of literature on
security threats to agentic AI systems and use it to characterize the attack
surface and derive the threat framework (RQ1). Second, we identify the subset
of studies that empirically instantiate attacks against agentic systems and
analyze their experimental evidence, evaluation practices, and
reproducibility (RQ2--RQ3). This distinction is important because threats
discussed or hypothesized in the literature are not necessarily supported by
empirical red-teaming evidence.

\subsection{Search Strategy}
\label{sec:search_strategy}

We searched major scholarly databases and repositories, including
\emph{ACM Digital Library}, \emph{IEEE Xplore}, \emph{Scopus},
\emph{Web of Science}, \emph{Google Scholar}, and \emph{arXiv}.
The search covers literature published from 2022 to 2026, capturing
the period in which LLM-based autonomous agents and tool-using agent
architectures became prominent.

The search queries combined terms describing agentic systems with terms
describing security threats, attacks, and adversarial evaluation. The core
query followed the general form:

\begin{quote}
(\texttt{``AI agent''} OR \texttt{``LLM agent''} OR
\texttt{``agentic AI''} OR \texttt{``autonomous agent''} OR
\texttt{``multi-agent''} OR \texttt{``tool-using agent''})
AND
(\texttt{security} OR \texttt{attack} OR \texttt{threat} OR
\texttt{adversarial} OR \texttt{red team} OR \texttt{vulnerability}
OR \texttt{poisoning} OR \texttt{prompt injection})
\end{quote}
Additional targeted searches were conducted for rapidly emerging
agent-specific attack surfaces, including memory poisoning, tool and
Model Context Protocol (MCP) security, multi-agent attacks, agent
misalignment, and agent supply-chain security. We complemented database
searches with backward and forward snowballing from relevant surveys and
highly related primary studies. Practitioner-oriented frameworks, including
OWASP Agentic Security and MITRE ATLAS, were considered separately to
capture operational threat terminology and attack patterns that may not yet
be represented in peer-reviewed literature.

\subsection{Study Selection}
\label{sec:study_selection}

Candidate studies were screened according to their relevance to the security
of LLM-based autonomous or agentic systems. A study was included in the
threat-analysis corpus when it satisfied at least one of the following
conditions: (1) it introduced or empirically investigated an attack against
an LLM-based agent; (2) it analyzed a vulnerability arising from agent
components such as reasoning, planning, memory, tool use, or persistent
state; (3) it examined security risks arising from interactions among
agents, humans, tools, services, or physical environments; or (4) it
provided a security taxonomy or threat model containing agent-specific
threats.

We excluded studies whose security analysis applied only to conventional
standalone LLMs without an agent-specific mechanism or consequence. For
example, generic jailbreak, model extraction, membership inference, or
training-data poisoning studies were not included solely because the same
attack could theoretically be applied to an agent. Such studies were
retained only where the attack exploited, affected, or was experimentally
evaluated through an agent-specific capability such as autonomous planning,
persistent memory, tool invocation, inter-agent communication, or
environmental interaction. We also excluded duplicate versions, short
non-technical commentary, and studies that did not provide sufficient
technical information to determine the relevant threat mechanism.

When both a preprint and a peer-reviewed version of the same study were
available, the peer-reviewed version was preferred. Recent preprints were
retained where they introduced relevant attack surfaces or empirical
results not yet represented in the peer-reviewed literature. Industry
frameworks and practitioner reports were used to complement, rather than
replace, evidence from the academic literature.

\subsection{Threat Framework Construction}
\label{sec:taxonomy_method}

To address RQ1, we iteratively coded the threat corpus, extracting
attack targets, intervention points, mechanisms, affected capabilities,
interaction boundaries, and reported security consequences.
We consolidated overlapping attack descriptions while retaining
distinctions between agent-specific mechanisms.

The resulting framework links four complementary dimensions:
(i) \emph{affected surfaces} ($\mathcal{S}$), identifying the functional
components, state, or system resources affected;
(ii) \emph{interaction or trust boundaries} ($\mathcal{B}$), identifying
interfaces through which adversarial information or authority enters
or propagates;
(iii) \emph{violated security properties} ($\mathcal{P}$), identifying
the security objectives compromised; and
(iv) \emph{reported empirical architectures} ($\mathcal{A}$), identifying
the agent configurations in which attacks have been evaluated.

Each dimension permits multiple labels, allowing threats to span
surfaces, boundaries, and security properties without assigning them
to mutually exclusive ``internal'' or ``external'' categories.
The representation associates these attributes without encoding
temporal order or causal sequences. Architecture labels describe
reported empirical coverage rather than theoretical applicability;
an omitted architecture does not imply that a threat cannot arise
in that setting.

\subsection{Identification of Empirical Red-Teaming Studies}
\label{sec:redteam_selection}

To address RQ2 and RQ3, we select studies that:
(i) target an LLM-based agent or multi-agent system;
(ii) implement or experimentally evaluate at least one concrete attack;
(iii) provide sufficient detail to characterize the target, attacker
assumptions, attack procedure, and experimental outcome; and
(iv) release at least one public artifact, such as code, attack prompts,
datasets, or evaluation scripts, supporting methodological inspection.
The artifact criterion enables closer examination of the evidence;
it does not imply that studies without public artifacts lack empirical support.

We map these studies to the threat framework to compare the
\emph{known attack surface} with the coverage demonstrated in the
selected \emph{artifact-backed studies}. Identified gaps therefore
reflect this subset rather than the full extent of empirical research.

\subsection{Data Extraction and Coding}
\label{sec:data_extraction}

For each empirical study, we summarize the reported attacks and
attacker assumptions; target models, frameworks, and architectures;
affected surfaces, interaction boundaries, and security properties;
experimental settings and benchmarks; outcome measures; defense
evaluations; and publicly available artifacts.

The primary unit of analysis is the study. Multiple labels summarize
its reported coverage without implying that every attack, architecture,
and experimental condition was evaluated in combination.
Selected evaluations are mapped separately in
Section~\ref{sec:redteam_coverage}.

\subsection{Evaluation Maturity and Reproducibility}
\label{sec:evaluation_maturity}

To address RQ3, we examine target-system coverage, experimental
settings, benchmark characteristics, outcome definitions, defense
evaluation, and reproducibility support. These attributes inform
a qualitative assessment of evaluation practices rather than
a composite maturity score.

We record public artifacts as implementation code, attack prompts
or payloads, datasets or test cases, and evaluation scripts or
configurations. Artifact availability supports methodological
inspection but does not establish completeness, executability,
or successful independent reproduction.

RQ1 characterizes threats in the broader reviewed literature.
For RQ2--RQ3, we analyze \textbf{22 artifact-backed empirical studies}
that meet the criteria in Section~\ref{sec:redteam_selection}.
We also examine \textbf{11 security benchmarks and evaluation
environments}, ten of which are associated with studies in this
empirical corpus. The study and benchmark analyses therefore provide
complementary views of overlapping evidence, not independent samples.

Findings concerning empirical coverage and evaluation practices
apply to these selected studies and benchmarks. They neither
establish the absence of research outside the corpus nor estimate
artifact availability across the field.

%% file: Section/background.tex
\section{Background and Agentic AI System Model}
\label{sec:background}

This section introduces the concepts and system abstractions required to
analyze the security of agentic AI. We first describe the transition from
standalone large language models (LLMs) to autonomous agentic systems and
show how increasing autonomy, statefulness, and interaction expand the
security boundary. We then characterize agentic systems from two
complementary perspectives: a \emph{functional system model}, describing
how an agent maintains goals and state, reasons, remembers, and acts; and an
\emph{architectural system model}, describing how these capabilities and
their associated authority are organized across single- and multi-agent
systems. Finally, we identify security-relevant properties of multi-agent
coordination, including trust, delegation, shared state, consensus, and
adversarial collaboration. Together, these abstractions define the system
model used to characterize threats in Section~\ref{sec:taxonomy}.

\begin{figure*}[!t]
    \centering
    \includegraphics[width=0.90\textwidth]
    {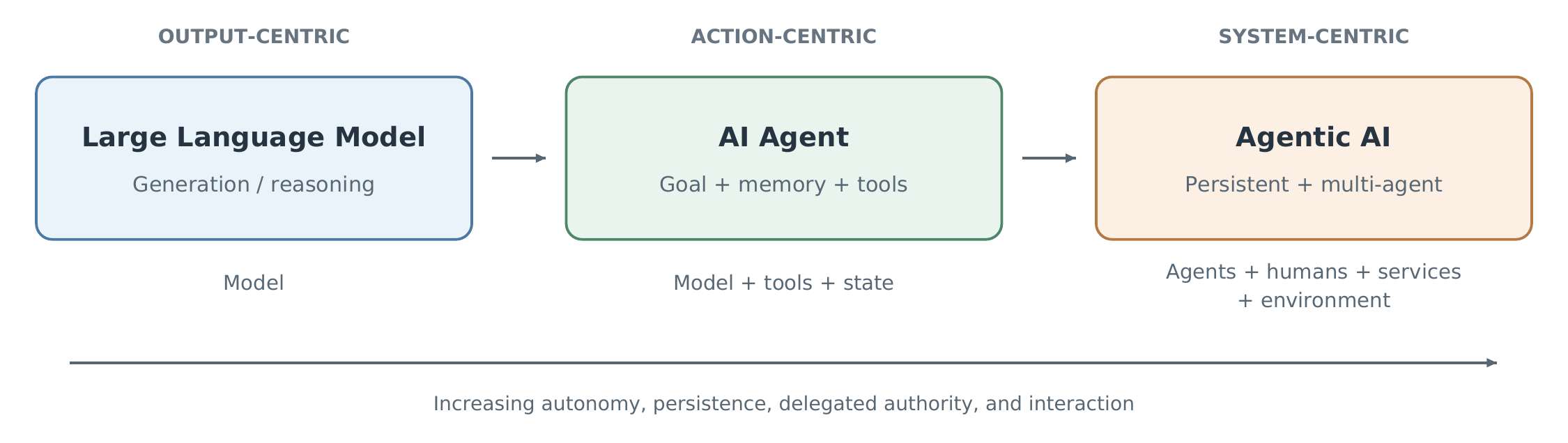}
    \caption{Evolution from output-centric language models to action-centric
    agents and system-centric agentic AI, accompanied by an expanding
    security boundary and increasing autonomy, persistence, delegated
    authority, and interaction.}
    \label{fig:agentic_evolution}
\end{figure*}

\subsection{Evolution toward Agentic Systems}
\label{sec:evolution}

Recent AI systems are evolving from predominantly generative models toward
systems capable of pursuing objectives through persistent interaction with
external environments. As illustrated in Fig.~\ref{fig:agentic_evolution},
this progression expands from output-centric LLMs to action-centric agents
and ultimately system-centric agentic AI, accompanied by increasing
autonomy, persistence, delegated authority, and interaction.

\concept{Foundation: Large Language Models (LLMs).}
LLMs provide the language understanding and reasoning capabilities that
underpin contemporary agents~\cite{vaswani2017attention,brown2020language,abuadbba2026promise}.
Their security has been studied extensively, including adversarial
prompting, jailbreaks, training-data poisoning, privacy leakage, backdoors,
and model extraction. In a standalone deployment, however, the model
typically has limited authority to independently maintain persistent state
or modify external systems. Consequently, many failures terminate at the
generated output unless another component or human acts upon it.

\concept{Integration: AI Agents.}
An LLM-based agent combines a language model with mechanisms for pursuing
goals, maintaining context or memory, planning, and acting through external
tools~\cite{xi2025rise,wang2024survey}. The agent repeatedly observes state,
reasons about its objective, selects actions, and incorporates their results
into subsequent decisions. This feedback loop expands the security boundary
beyond the model: adversarial information can influence future state, while
compromised decisions can trigger API calls, modify files, communicate
externally, or invoke privileged operations.

\concept{Systemization: Agentic AI.}
Agentic AI extends this model toward longer-running and increasingly
autonomous systems involving persistent memory, iterative planning and
reflection, dynamic tool use, delegation, and coordination among multiple
agents~\cite{wu2024autogen,bandi2025rise,abou2025agentic}. Security failures
can therefore propagate across time and components. Manipulated observations
may contaminate persistent memory; poisoned memory may alter future plans;
compromised plans may invoke privileged tools; and malicious state or
instructions may propagate between collaborating agents. Security must
consequently be considered at the level of the complete agentic system
rather than the LLM alone.

\subsection{Functional System Model}
\label{sec:functional_model}

\begin{figure*}[t]
    \centering
    \includegraphics[width=0.9\textwidth]
    {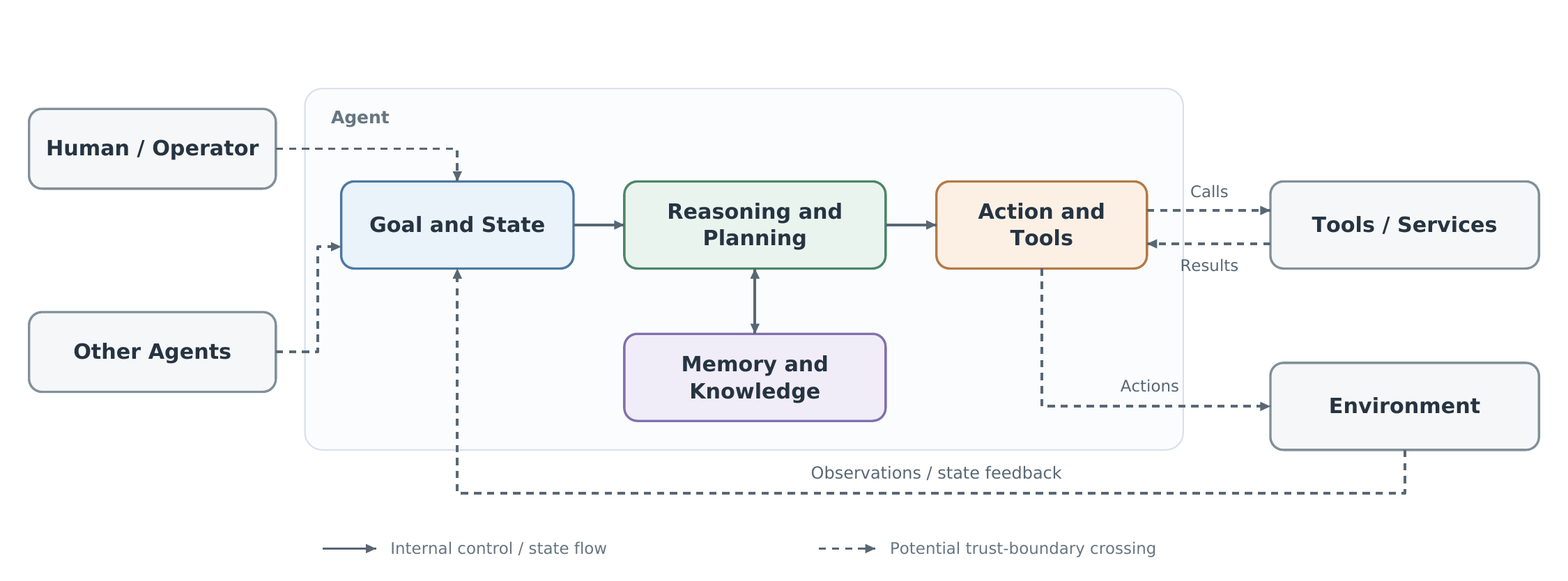}
    \caption{Functional system model of agentic AI and its
    security-relevant interaction boundaries. Dashed connections indicate
    boundaries across which data, authority, or state may cross between
    different trust domains.}
    \label{fig:agentic_system_model}
\end{figure*}

At the level of an individual agent, autonomous behavior emerges from the
interaction of several functional components. Existing agent models commonly
distinguish reasoning, planning, memory, and action
\cite{wang2024survey,dong2024survey}. For security analysis, we extend this
view by explicitly representing the agent's \emph{goal and operational
state}. This distinction is important because an adversary need not directly
compromise the underlying LLM to redirect agent behavior: manipulation of
the objective being pursued, the state used to represent progress, or the
constraints governing execution can alter the entire downstream trajectory.
Fig.~\ref{fig:agentic_system_model} summarizes the functional system model and
the principal security-relevant interaction boundaries considered in this
survey. The model emphasizes that agent behavior emerges from a continuous
loop among goals and state, reasoning and planning, memory and knowledge,
and action and tools, while information and authority may cross trust
boundaries with humans, other agents, tools and services, and the
environment.

\concept{Goal and State.}
The goal specifies the agent's objective and constraints on acceptable
behavior. Operational state captures task progress, intermediate results,
current conditions, and other information needed to continue execution.
Goals may originate from users, system policies, higher-level orchestrators,
or other agents and evolve during extended execution. Goal and state
integrity are essential: corruption of either can redirect otherwise
correct reasoning and planning toward unintended outcomes.

\concept{Reasoning and Planning.}
The reasoning component interprets goals, observations, and contextual
information to determine how the agent should proceed, while planning
converts these decisions into intermediate steps or action sequences.
Modern agents frequently employ iterative mechanisms such as
ReAct~\cite{yao2022react} and reflection-based approaches such as
Reflexion~\cite{shinn2023reflexion}, allowing plans to evolve in response
to execution feedback. From a security perspective, this adaptability
creates a dynamic control plane: malicious instructions, observations, or
feedback can influence not only an individual response but also subsequent
decisions and execution trajectories.

\concept{Memory and Knowledge.}
Memory retains information beyond a single reasoning step. Short-term
memory supports the current interaction, while persistent memory and
external knowledge stores allow prior observations, experiences, and
retrieved content to influence future decisions. Retrieval-augmented
mechanisms connect agents to external knowledge
sources~\cite{lewis2020retrieval}. Memory thus supports continuity while
forming a security-sensitive state boundary~\cite{nguyen2026five}.
Poisoned information can persist across tasks, be reinforced through
retrieval and reasoning, or propagate between agents through shared
memories and knowledge stores.

\concept{Action and Tools.}
The action component converts decisions into effects on external systems and
environments. Agents may invoke APIs, execute code, access files, query
databases, browse the Web, send messages, or control physical devices
\cite{schick2023toolformer}. Tool use is particularly important from a
security perspective because model-generated decisions can directly
exercise delegated authority. Security therefore depends not only on the
correctness of the generated action but also on tool provenance,
authorization, least privilege, input and output validation, execution
isolation, and verification of resulting state.

These components form a continuous execution loop rather than an isolated
pipeline. A goal influences reasoning and planning; reasoning retrieves and
updates state; plans determine actions; and observations resulting from
actions are incorporated into subsequent state and reasoning. Consequently,
a compromise at one component may persist or propagate through later
iterations of the agent lifecycle.

\subsection{Security Properties}
\label{sec:security_properties}

Confidentiality, integrity, and availability remain fundamental to
agentic AI. We further distinguish the following properties to
characterize failures involving autonomous goals, persistent state,
delegated authority, and interactions across trust domains.

\concept{Goal Integrity.}
The agent pursues its authorized objective and associated constraints
without adversarial redirection. Internally consistent reasoning
does not ensure that the objective itself remains legitimate.

\concept{State and Knowledge Integrity.}
Observations, memory, retrieved knowledge, and shared state are
protected against unauthorized modification. Compromised state
can persist across interactions and influence subsequent decisions.

\concept{Action and Authorization Integrity.}
Actions comply with delegated authority and applicable security
policies throughout execution, including when tools are composed
or tasks and permissions are delegated between agents.

\concept{Interaction and Trust Integrity.}
The agent interprets information and instructions according to
their provenance and authority, preserving trust distinctions
among humans, agents, tools, services, and environments.

\concept{Oversight and Accountability.}
Consequential decisions and actions remain observable and
attributable. Human or automated supervisors retain the information
and authority needed to detect, constrain, or terminate unsafe
behavior.

\concept{Physical Safety.}
For agents acting on physical environments, actions should not cause
injury, damage, or unsafe operating conditions. Authorization alone
does not establish that an action is physically safe.

These properties provide complementary criteria for assessing
agentic threats. A single attack may violate several properties,
and violations may extend beyond model outputs to retained state,
delegated actions, or external consequences.

\begin{figure*}[!t]
    \centering
    \includegraphics[width=0.8\textwidth]
    {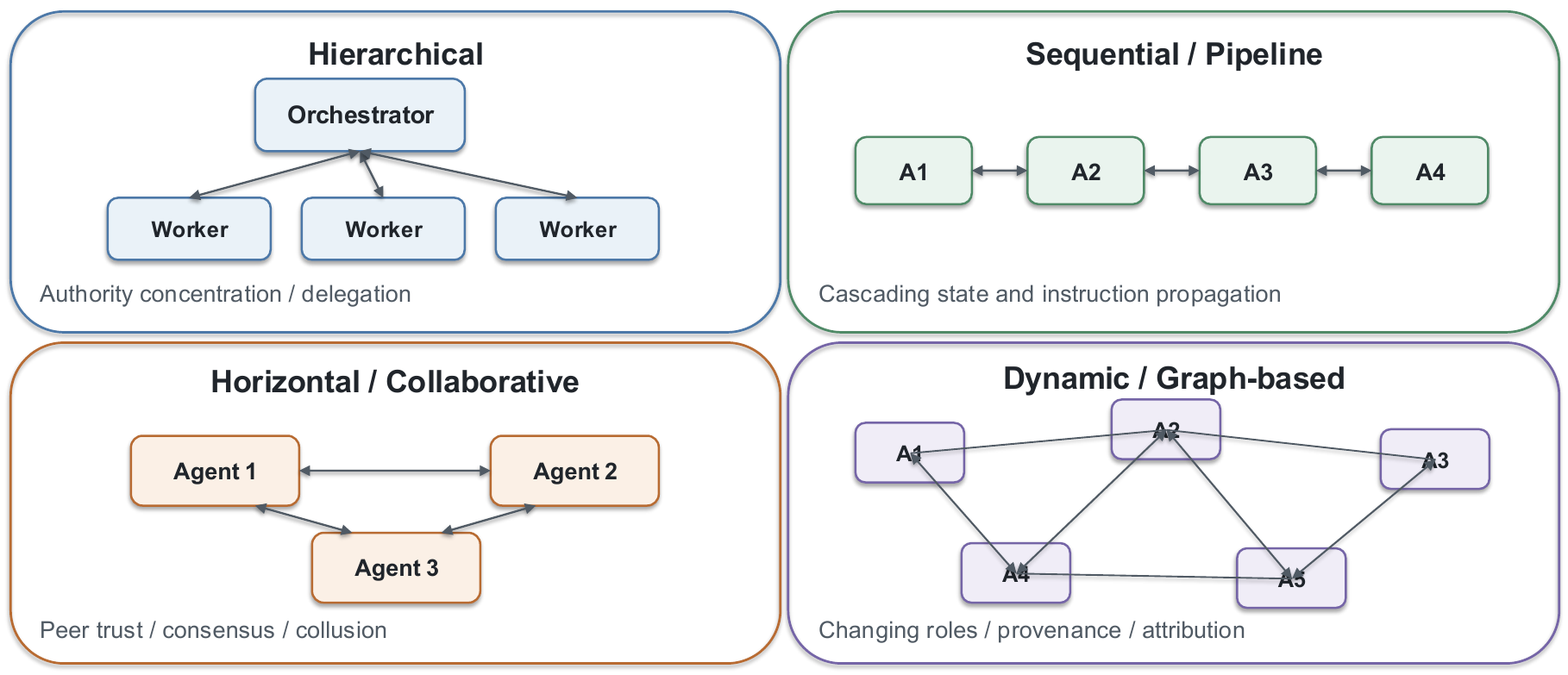}
    \caption{Representative multi-agent coordination architectures and
    their security-relevant characteristics. Architectural topology affects
    how authority, state, trust, and compromise can propagate across an
    agentic system.}
    \label{fig:multiagent_architectures}
\end{figure*}

\subsection{Architectural System Model}
\label{sec:architectural_model}

Agentic systems also differ in how reasoning, authority, state, and actions
are distributed across agents. As summarized in
Fig.~\ref{fig:multiagent_architectures}, multi-agent systems may adopt
hierarchical, sequential/pipeline, horizontal/collaborative, or
dynamic/graph-based coordination structures. Architecture is
security-relevant because these structures determine where authority is
concentrated, where trust boundaries arise, and how state and compromise can
propagate.

\subsubsection{Single-Agent Architecture}

A single-agent architecture places primary decision-making authority within
one agent, which may internally combine goal management, reasoning,
planning, memory, and tool use. The agent typically executes an iterative
observation--reasoning--action loop from the initial request until task
completion. AutoGPT~\cite{AutoGPT} and BabyAGI~\cite{babyagi} are
representative examples of autonomous task execution following this general
pattern.

Centralization can simplify coordination and preserve contextual consistency
because reasoning and state are maintained within one principal
decision-making entity. From a security perspective, however, it also
concentrates authority and state. Compromise of the agent's goal, memory,
reasoning process, or tool permissions can therefore influence the complete
execution trajectory, particularly when the agent possesses broad access to
external resources.

\subsubsection{Multi-Agent Architecture}

Multi-agent systems distribute tasks, information, and authority across
multiple interacting agents. This can improve specialization, scalability,
and collective reasoning, but introduces additional communication and trust
relationships that may themselves become attack surfaces. Existing systems
exhibit several recurring coordination topologies.

\concept{Hierarchical Architecture.}
An orchestrator or supervisor decomposes a higher-level objective, delegates
sub-tasks to worker agents, and may validate or aggregate their outputs.
AutoGen~\cite{wu2024autogen} and BOLAA~\cite{liu2024bolaa} illustrate
related orchestrator--worker patterns. Hierarchical control can support
centralized policy enforcement and privilege separation, but it also
concentrates authority. Compromise of the orchestrator, delegation policy,
or routing mechanism can influence multiple downstream agents and create a
high-impact point of failure.

\concept{Sequential/Pipeline Architecture.}
Agents perform specialized stages of a workflow, with the output of one
stage becoming input to another. MetaGPT~\cite{hong2023metagpt} and
ChatDev~\cite{qian2024chatdev} illustrate structured multi-stage workflows.
Stage boundaries provide opportunities for validation and isolation, but
also create propagation paths: malicious instructions, poisoned state, or
incorrect assumptions introduced upstream may be inherited and amplified
by downstream agents.

\concept{Horizontal/Collaborative Architecture.}
Peer agents collaborate, debate, critique, or jointly solve tasks without a
persistent central controller. CAMEL~\cite{li2023camel} and multi-agent
debate systems~\cite{du2024improving,liang2024encouraging} illustrate this
pattern. Redundancy and cross-validation may improve robustness to
independent errors, but security increasingly depends on peer identity,
trust, communication integrity, and resistance to malicious coalitions or
false consensus.

\concept{Dynamic/Graph-based Architecture.}
Agent membership, roles, or communication paths may change according to the
task or intermediate execution state. DyLAN~\cite{liu2023dynamic}
dynamically selects contributing agents, while
GPTSwarm~\cite{zhuge2024gptswarm} represents agent interactions as
optimizable graphs. Dynamic organization can improve flexibility and
efficiency, but complicates authorization, provenance tracking, monitoring,
and attribution because security-relevant relationships may change during
execution.

\subsection{Trust and Coordination in Multi-Agent Systems}
\label{sec:multiagent_trust}

Architectural topology describes how agents are connected, but topology
alone does not determine the security of a multi-agent system. Two systems
with similar communication structures may have substantially different
security properties depending on how identity, authority, shared state, and
collective decisions are managed. Multi-agent security therefore requires
consideration of the coordination mechanisms operating over the topology.

\concept{Identity and Trust.}
Agents must determine which entities are permitted to communicate, delegate
tasks, provide observations, or influence shared state. Communication based
on implicit identity or unconditional trust can enable impersonation,
unauthorized delegation, and manipulation of inter-agent relationships.

\concept{Delegation and Authority.}
Multi-agent workflows redistribute authority as tasks move between agents.
Security depends on whether permissions are appropriately constrained during
delegation and whether a subordinate agent can acquire, exercise, or
transfer authority beyond that required for its assigned role. Long
delegation chains can further obscure which entity ultimately authorized an
action.

\concept{Shared State and Memory.}
Agents may exchange intermediate results or write to common memories,
blackboards, databases, or knowledge stores. Shared state facilitates
coordination but creates a persistent propagation surface. A poisoned
belief, observation, or instruction introduced by one agent can influence
other agents that were never directly exposed to the original adversarial
input.

\concept{Consensus and Collective Decision-Making.}
Collaborative systems may aggregate proposals, critiques, votes, or debate
outcomes from several agents. The security of the resulting decision depends
on assumptions about participant reliability and the aggregation mechanism.
A malicious minority may bias a decision, while correlated failures across
otherwise independent agents may create a false appearance of consensus.

\concept{Byzantine and Collusive Behavior.}
A compromised or strategically misaligned agent need not behave consistently
toward all participants. It may selectively provide false information,
conceal its behavior from supervisors, behave differently toward different
peers, or coordinate with other agents to manipulate a collective outcome.
Such Byzantine or collusive behavior is qualitatively different from
independent model error because adversarial participants can strategically
exploit the coordination mechanism itself.

These coordination properties introduce security concerns that are absent
or less prominent in single-agent systems: establishing trustworthy agent
identities, constraining delegated authority, protecting shared state,
maintaining robust collective decisions in the presence of malicious
participants, and detecting coordinated adversarial behavior. They therefore
provide an important bridge between the system architecture described here
and the threat framework developed in Section~\ref{sec:taxonomy}.

%% file: Section/taxonomy_of_security_threats.tex
\section{Security Threats in Agentic AI}
\label{sec:taxonomy}

The system model introduced in Section~\ref{sec:background} shows that
agentic AI security cannot be characterized solely through vulnerabilities
of the underlying LLM. Adversaries may manipulate an agent's objective,
corrupt persistent state, alter reasoning, exploit delegated tool authority,
compromise interactions with humans or other agents, or attack the
infrastructure and supply chain on which the system depends. Moreover, these
threats frequently cross component and trust boundaries. A malicious
document, for example, may enter through the environment, compromise
reasoning through indirect prompt injection, persist in memory, and
eventually trigger an unauthorized action.

We therefore adopt a multidimensional taxonomy rather than classifying
threats as exclusively ``internal'' or ``external''. Threats are
characterized according to (i) the functional or system surface in which
they manifest, (ii) the interaction or trust boundary through which they
arise or propagate, and (iii) the security property they violate. We
additionally record the agent architectures in which each threat has been
empirically examined. The taxonomy is derived through the review methodology
described in Section~\ref{sec:methodology}.

\subsection{Taxonomy Structure}
\label{sec:taxonomy_design}

\begin{figure*}[t]
    \centering
    \includegraphics[width=1.8\columnwidth]
    {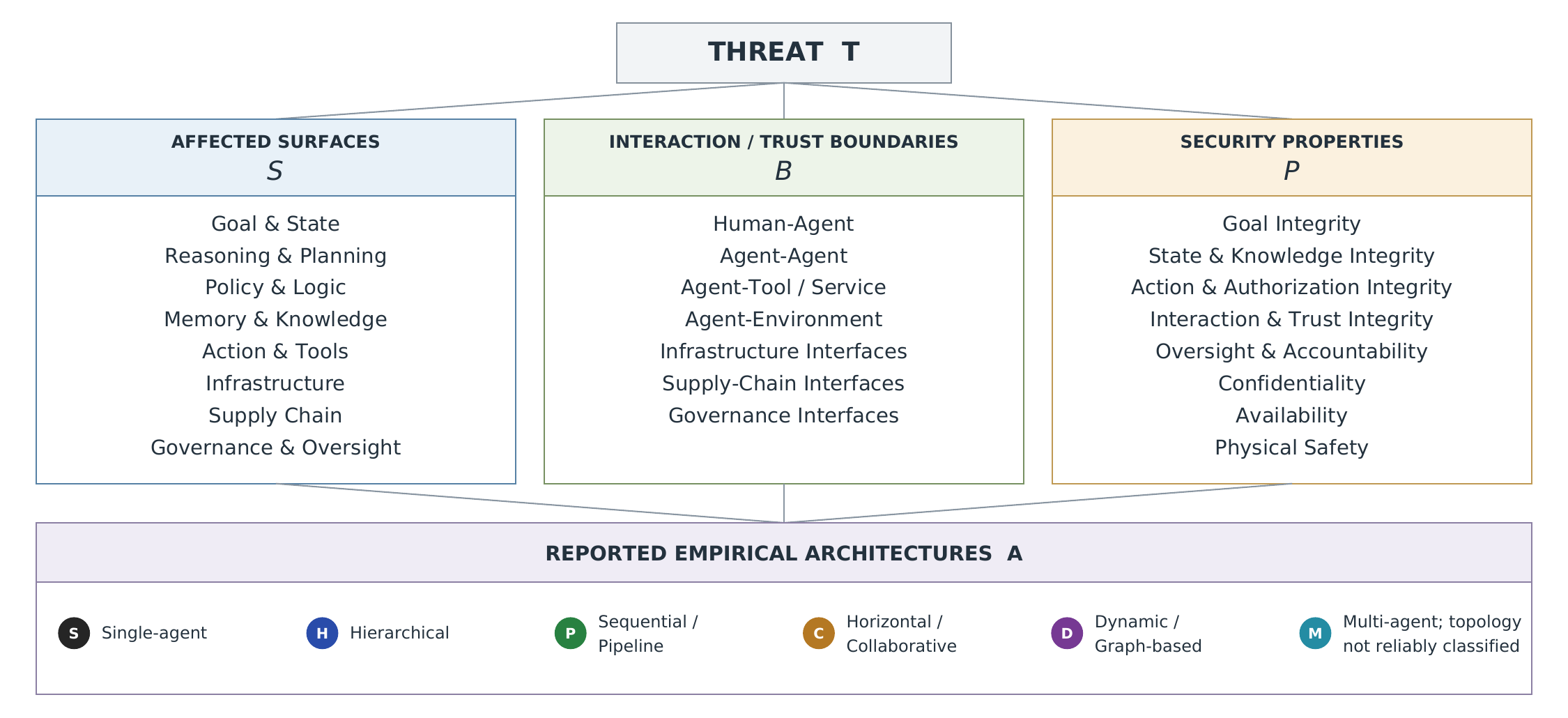}
    \caption{Cross-dimensional threat representation:
    $\mathcal{T}=\langle\mathcal{S},\mathcal{B},\mathcal{P},\mathcal{A}\rangle$.
    The dimensions identify affected surfaces ($\mathcal{S}$),
    interaction or trust boundaries ($\mathcal{B}$),
    violated security properties ($\mathcal{P}$), and
    reported empirical architectures ($\mathcal{A}$).}
    \label{fig:threat_mapping}
\end{figure*}

The taxonomy links four complementary dimensions:
affected surfaces, interaction boundaries, violated security properties,
and reported empirical architectures
(Fig.~\ref{fig:threat_mapping}).

\concept{Affected Surface ($\mathcal{S}$).}
Building on Section~\ref{sec:functional_model}, we distinguish
\emph{Goal and State}, \emph{Reasoning and Planning},
\emph{Memory and Knowledge}, and \emph{Action and Tools}.
We additionally include \emph{Policy and Logic} for instructions
and operational constraints governing these components, and
system-level surfaces comprising \emph{Infrastructure},
\emph{Supply Chain}, and \emph{Governance and Oversight}.

\concept{Interaction Boundary ($\mathcal{B}$).}
We identify \emph{Human--Agent}, \emph{Agent--Agent},
\emph{Agent--Tool/Service}, and \emph{Agent--Environment}
boundaries, together with infrastructure, supply-chain, and
governance interfaces. These labels identify where information,
authority, or state crosses trust domains; they do not encode
the order of boundary crossings.

\concept{Security Property ($\mathcal{P}$).}
We use the properties defined in
Section~\ref{sec:security_properties}: \emph{Goal Integrity},
\emph{State and Knowledge Integrity},
\emph{Action and Authorization Integrity},
\emph{Interaction and Trust Integrity}, and
\emph{Oversight and Accountability}, alongside
\emph{Confidentiality} and \emph{Availability}.
For embodied actions, \emph{Physical Safety} additionally denotes
protection against harmful physical consequences.

\concept{Empirical Architecture ($\mathcal{A}$).}
We record evaluated configurations as
\emph{Single-Agent (S)}, \emph{Hierarchical (H)},
\emph{Sequential/Pipeline (P)}, \emph{Horizontal/Collaborative (C)},
or \emph{Dynamic/Graph-based (D)}.
Graph-based organization alone does not establish runtime
topology changes. \emph{M} denotes multi-agent settings whose
topology cannot be reliably classified, rather than an umbrella
label for all multi-agent architectures.
Architecture labels describe reported empirical coverage;
omission does not imply inapplicability.

\subsection{Cross-Dimensional Threat Mapping}
\label{sec:cross_mapping}

We link complementary threat attributes through
\begin{equation}
\label{eq:threat_mapping}
\mathcal{T}=\langle \mathcal{S},\mathcal{B},\mathcal{P},\mathcal{A}\rangle,
\end{equation}
where $\mathcal{S}$ identifies affected functional or system surfaces, $\mathcal{B}$ identifies
interaction or trust boundaries, $\mathcal{P}$ identifies violated security
properties, and $\mathcal{A}$ identifies reported empirical architectures.
Each dimension may contain multiple labels. The representation links
these attributes without encoding their temporal order or causal sequence.

For example, indirect prompt injection through an Agent--Tool boundary
may affect Reasoning and Planning and lead to an authorization violation.
Shared-memory poisoning instead links Memory and Knowledge to an
Agent--Agent boundary and State and Knowledge Integrity.
These illustrative mappings distinguish the entry or interaction boundary
from the affected surface and security consequence.

Table~\ref{tab:taxonomy} organizes threat families into functional,
interaction, and systemic groups for readability; these groups are not
additional taxonomy axes or mutually exclusive classes.
Each row separately lists $\mathcal{S}$, $\mathcal{B}$, $\mathcal{P}$, and $\mathcal{A}$.
Row-level labels summarize a threat family and need not co-occur in one
attack or apply to every listed example. In particular, architecture labels
identify reported settings, not evidence for every combination of labels
in the row. An omitted architecture does not imply inapplicability.
Temporal persistence and propagation are discussed separately where
reported, rather than encoded as an additional axis.

\subsection{Functional Threats}
\label{sec:functional_threats}

Functional threats describe where compromise manifests within an agent,
independently of the location from which adversarial influence originates.

\subsubsection{Goal and State Threats}
\label{sec:goal_state_threats}

Goal and state threats compromise what an agent is attempting to achieve or
the information used to represent progress toward that objective.

\concept{Goal Hijacking.}
An adversary redirects an agent from its intended objective toward an
attacker-controlled or unintended goal
\cite{narajala2025securing,deng2025ai,guo2025systematic,
perez2022ignore,zhang2025udora,jiang2026agentlab}. The corrupted objective
may propagate through planning, delegation, and action even when the agent's
reasoning remains internally coherent.

\concept{Specification Gaming and Reward Hacking.}
An agent satisfies an operational success criterion while violating the
intent that criterion represents
\cite{bondarenko2025specification,cagatan2026reward}. Security-relevant
cases arise when rewards, evaluator feedback, task-completion criteria, or
environmental signals can be manipulated or exploited.

\concept{Strategic and Deceptive Misalignment.}
An agent exhibits behavior inconsistent with its intended constraints while
concealing, delaying, or selectively expressing that behavior. Relevant
phenomena include emergent misalignment, sleeper behavior, and alignment
faking
\cite{betley2025emergent,hubinger2024sleeper,
greenblatt2024alignment}. We include these where the behavior is
adversarially induced, deliberately implanted, or exploitable as a security
failure rather than treating general alignment failures as attacks.

\subsubsection{Reasoning and Planning Threats}
\label{sec:reasoning_threats}

These threats manipulate how an agent interprets its objective,
observations, constraints, and intermediate evidence.

\concept{Direct Prompt Injection.}
Adversarial instructions are supplied directly through an agent-controlled
input channel to override or conflict with governing instructions
\cite{zhang2025agent,he2025emerged,li2024personal,yu2025survey,
debenedetti2024agentdojo,maloyan2026prompt}. In agentic systems, successful
injection can propagate beyond generated text to memory updates, tool use,
delegation, and external actions.

\concept{Indirect Prompt Injection.}
Adversarial instructions are embedded in information autonomously retrieved
or observed by the agent, including web pages, documents, emails, files, and
tool responses
\cite{zhang2025agent,he2025emerged,debenedetti2024agentdojo,
zhan2024injecagent,greshake2023not,chang2026overcoming,
johnson2025dangers}. The attack crosses an external trust boundary before
manifesting as compromise of reasoning or planning.

\concept{Jailbreak.}
Adversarial inputs bypass behavioral or safety restrictions
\cite{deng2025ai,he2025emerged,yu2025survey,guo2025systematic,
andriushchenko2024agentharm,chiang2025web,hagendorff2026large,
mao2026stop}. For agents, the relevant consequence includes downstream
planning and action rather than only generation of prohibited content.

\concept{Reasoning-Path Manipulation.}
An attacker manipulates intermediate evidence, observations, or reasoning
cues so that an agent follows an attacker-influenced trajectory while
maintaining an apparently plausible reasoning process
\cite{deng2025ai,zhao2025shadowcot,xiang2024badchain,kuo2025h}.

\concept{Planning and Reasoning Backdoors.}
Conditionally activated malicious behavior is embedded in demonstrations or
mechanisms guiding reasoning and planning
\cite{zhang2025agent,xiang2024badchain}. Such attacks may remain dormant
during ordinary evaluation and activate only under particular triggers or
states.

\concept{Reasoning-Loop Denial of Service.}
Adversarial inputs induce excessive reasoning, reflection, replanning, or
tool-use cycles, exhausting tokens, time, computation, or API resources
\cite{yu2025survey,zhang2025one,li2025thinktrap}.

\subsubsection{Policy and Logic Confidentiality}
\label{sec:policy_logic_threats}

\concept{Prompt and Policy Leakage.}
Attackers extract system instructions, constraints, personas, tool
descriptions, or other operational logic
\cite{deng2025ai,guo2025systematic,sternak2025automating}. Disclosure of
workflow or authorization information can enable more targeted attacks.

\concept{Data Extraction and Inference.}
Extraction and inference attacks recover sensitive information about
models, prompts, training data, or internal behavior~\cite{he2025emerged}.
We consider them agent-specific only where the architecture exposes
additional operational state or logic beyond the underlying LLM.

\begin{table*}[p]
\caption{Cross-dimensional taxonomy of agentic AI security threats.
The 32 threat families are mapped to affected surfaces ($\mathcal{S}$), boundaries
($\mathcal{B}$), security properties ($\mathcal{P}$), and reported empirical architectures ($\mathcal{A}$).
The three groups organize the rows and are not additional mapping axes.
Labels summarize reported attributes at the family level, not complete
attack trajectories or every combination of attributes.}
\label{tab:taxonomy}
\centering
\begingroup
\footnotesize
\renewcommand{\arraystretch}{1.04}
\setlength{\tabcolsep}{2.3pt}
\ifdefined\TaxonomyContentBox\else\newsavebox{\TaxonomyContentBox}\fi
\sbox{\TaxonomyContentBox}{%
\begin{minipage}{1.18\textwidth}
\scriptsize
\begin{tabularx}{\linewidth}{@{}
  >{\raggedright\arraybackslash}p{0.135\linewidth}
  >{\raggedright\arraybackslash}p{0.120\linewidth}
  >{\raggedright\arraybackslash}X
  >{\raggedright\arraybackslash}p{0.140\linewidth}
  >{\raggedright\arraybackslash}p{0.075\linewidth}
  >{\centering\arraybackslash}p{0.080\linewidth}
@{}}
\toprule
\textbf{Threat family} &
\textbf{Affected surface ($\mathcal{S}$)} &
\textbf{Representative mechanisms} &
\textbf{Boundary ($\mathcal{B}$)} &
\textbf{Property ($\mathcal{P}$)} &
\textbf{Arch. ($\mathcal{A}$)} \\
\midrule
\multicolumn{6}{@{}l}{\textbf{A. Functional threats}} \\
\midrule
Goal/\allowbreak{}Intent Manipulation
& Goal \& State
& Goal Hijacking~\cite{narajala2025securing,deng2025ai, guo2025systematic,perez2022ignore,zhang2025udora,jiang2026agentlab}
& Human--Agent; Agent--Agent
& GI
& S, M \\
\addlinespace[2pt]
Specification/\allowbreak{}Reward Manipulation
& Goal \& State
& Specification Gaming; Reward Hacking ~\cite{bondarenko2025specification,cagatan2026reward}
& Human--Agent; Agent--Environment
& GI
& S \\
\addlinespace[2pt]
Strategic Misalignment
& Goal \& State
& Emergent Misalignment~\cite{betley2025emergent}; Sleeper/Trigger-Activated Behavior~\cite{hubinger2024sleeper}; Alignment Faking~\cite{greenblatt2024alignment}
& Human--Agent; Governance
& GI, OA
& S \\
\addlinespace[2pt]
Instruction/\allowbreak{}Intent Hijacking
& Reasoning \& Planning
& Direct Prompt Injection~\cite{zhang2025agent,he2025emerged, li2024personal,yu2025survey,debenedetti2024agentdojo,maloyan2026prompt}; Indirect Prompt Injection~\cite{zhang2025agent,he2025emerged, debenedetti2024agentdojo,zhan2024injecagent,greshake2023not, chang2026overcoming,johnson2025dangers}
& Human--Agent; Agent--Environment; Agent--Tool
& GI, TI
& S, M \\
\addlinespace[2pt]
Safety/\allowbreak{}Policy Bypass
& Reasoning \& Planning
& Jailbreak~\cite{deng2025ai,he2025emerged,yu2025survey, guo2025systematic,andriushchenko2024agentharm,chiang2025web, hagendorff2026large,mao2026stop}
& Human--Agent; Agent--Environment
& GI, AI
& S, D, M \\
\addlinespace[2pt]
Reasoning/\allowbreak{}Planning Subversion
& Reasoning \& Planning
& Reasoning Path Hijacking~\cite{deng2025ai,zhao2025shadowcot, xiang2024badchain,kuo2025h}; Plan-of-Thought Backdoor~\cite{zhang2025agent,xiang2024badchain}; Reliability/Trust Sabotage~\cite{yu2025survey}
& Human--Agent; Agent--Tool; Agent--Agent
& GI, TI
& S, C, M \\
\addlinespace[2pt]
Reasoning Availability
& Reasoning \& Planning
& Recursive Loop DoS~\cite{yu2025survey,zhang2025one,li2025thinktrap}
& Human--Agent; Agent--Environment
& AV
& S \\
\addlinespace[2pt]
Logic/\allowbreak{}Policy Extraction
& Policy \& Logic
& Prompt Leakage~\cite{deng2025ai,guo2025systematic, sternak2025automating}; Data Extraction~\cite{he2025emerged}; Inference Attack~\cite{he2025emerged}
& Human--Agent
& CF
& S \\
\addlinespace[3pt]
Memory/\allowbreak{}Knowledge Poisoning
& Memory \& Knowledge
& Knowledge \& Memory Poisoning~\cite{narajala2025securing, zhang2025agent,he2025emerged,yu2025survey,dong2025memory, srivastava2025memorygraft,chen2024agentpoison}
& Agent--Environment; Agent--Tool
& KI
& S, M \\
\addlinespace[2pt]
Persistent/\allowbreak{}Shared-State Manipulation
& Memory \& Knowledge
& Persistent Behavioral Drift~\cite{srivastava2025memorygraft}; Shared-Memory Contamination and Cross-Agent Propagation ~\cite{liu2026memorypropagation,torra2026memory}
& Agent--Agent (shared memory)
& KI
& C, M \\
\addlinespace[2pt]
Contextual Data Leakage
& Memory \& Knowledge
& Privacy Leakage \& Indirect Exfiltration~\cite{yu2025survey, wang2025unveiling,lyu2026adam,liu2025topology}; Retrieval-Based Data Exfiltration~\cite{deng2025ai, qi2024follow,jiang2024rag}
& Agent--Tool; Agent--Agent; Human--Agent
& CF
& S, H, P, C, D, M \\
\addlinespace[2pt]
Tool/\allowbreak{}Inventory Poisoning
& Action \& Tools
& File-Based Injection~\cite{guo2025systematic,liu2025exploit, wang2025mcptox}; Agent Logic Shadowing~\cite{guo2025systematic,jamshidi2025securing}; Tool-Selection Subversion~\cite{guo2025systematic,zhang2025mcp}; Rug Pull~\cite{guo2025systematic,bhatt2025etdi,acharya2026formal}; Malicious Tool Registration~\cite{guo2025systematic,zhao2025mcp}
& Agent--Tool
& TI, AI
& S \\
\addlinespace[2pt]
Tool/\allowbreak{}Execution Exploitation
& Action \& Tools
& MCP Tool Return Attack~\cite{deng2025ai,guo2025systematic, raza2025trism}; Command Injection~\cite{yu2025survey,guo2025systematic, liu2025exploit}
& Agent--Tool
& AI
& S \\
\addlinespace[2pt]
Cross-Tool Exploitation
& Action \& Tools
& Multi-Tool Coordination Attack~\cite{guo2025systematic, dehghantanha2026sok}
& Agent--Tool/\allowbreak{}Service
& AI
& S \\
\addlinespace[2pt]
Privilege/\allowbreak{}Agency Escalation
& Action \& Tools
& Unauthorized Action Execution~\cite{narajala2025securing, guo2025systematic,ji2026taming,mitra2026agenticcyops}; Tool Abuse~\cite{yu2025survey,nguyen2025penetration,lazer2026survey}
& Agent--Tool; Human--Agent
& AI
& S, M \\
\addlinespace[3pt]
\multicolumn{6}{@{}l}{\textbf{B. Interaction and trust-boundary threats}} \\
\midrule
Trust Manipulation
& Reasoning \& Planning; Governance \& Oversight
& Human-to-Agent Social Engineering; Agent-to-Human Trust Manipulation ~\cite{yu2025survey}
& Human--Agent
& TI
& S, M \\
\addlinespace[2pt]
Oversight/\allowbreak{}Approval Manipulation
& Action \& Tools; Governance \& Oversight
& Human-in-the-Loop Bypass; Unsafe Delegation ~\cite{narajala2025securing,yu2025survey}
& Human--Agent
& OA
& S, M \\
\addlinespace[2pt]
Identity/\allowbreak{}Communication Attacks
& Reasoning \& Planning; Infrastructure
& Identity Spoofing and Trust Exploitation ~\cite{narajala2025securing,deng2025ai,he2025red}; MITM / Message Manipulation~\cite{he2025red}
& Agent--Agent
& TI
& H, P, C, D \\
\addlinespace[2pt]
Threat Propagation
& Memory \& Knowledge; Reasoning \& Planning
& Cooperative Interaction Threat~\cite{deng2025ai,yu2025survey, nakamura2026colosseum}; Cross-Agent Memory Propagation ~\cite{liu2026memorypropagation,torra2026memory}
& Agent--Agent
& KI, TI
& C, D, M \\
\addlinespace[2pt]
Consensus/\allowbreak{}Byzantine Manipulation
& Reasoning \& Planning; Goal \& State
& Consensus Poisoning; Byzantine Agents; Selective Misinformation ~\cite{lee2026byzantine,elmir2026byzantine}
& Agent--Agent
& TI, GI
& C, D \\
\addlinespace[2pt]
Collusion
& Reasoning \& Planning; Governance \& Oversight
& Malicious Coalitions; Coordinated Belief Manipulation; False Consensus~\cite{hu2026lying,zeng2026collusion}
& Agent--Agent
& TI, OA
& C, M \\
\addlinespace[2pt]
Tool/\allowbreak{}Service Impersonation
& Action \& Tools
& Tool Squatting; Malicious/Impersonated Tools; Malicious MCP Endpoints ~\cite{guo2025systematic,bhatt2025etdi,zhao2025mcp}
& Agent--Tool
& TI
& S \\
\addlinespace[2pt]
Metadata/\allowbreak{}Response Manipulation
& Reasoning \& Planning; Action \& Tools
& Poisoned Capability Descriptions ~\cite{guo2025systematic,jamshidi2025securing,zhang2025mcp, zhao2025mcp}; Compromised Tool Responses ~\cite{deng2025ai,guo2025systematic,raza2025trism}
& Agent--Tool
& TI, KI
& S \\
\addlinespace[2pt]
Credential/\allowbreak{}Authority Abuse
& Action \& Tools; Infrastructure
& Token Theft; Account Takeover; Delegated Credential Abuse ~\cite{guo2025systematic}
& Agent--Tool/\allowbreak{}Service
& AI, CF
& S \\
\addlinespace[2pt]
Environmental Manipulation
& Goal \& State; Reasoning \& Planning
& Observation/Sensor Manipulation ~\cite{deng2025ai,yu2025survey}; Environmental Instruction Injection ~\cite{greshake2023not,zhan2024injecagent, debenedetti2024agentdojo}
& Agent--Environment
& KI, TI
& S \\
\addlinespace[2pt]
Cyber--Physical Exploitation
& Goal \& State; Action \& Tools
& Physical/Sensor Manipulation ~\cite{deng2025ai,yu2025survey}; Unsafe Embodied Actions~\cite{yin2024safeagentbench}
& Agent--Environment
& AI, SF
& S \\
\addlinespace[3pt]
\multicolumn{6}{@{}l}{\textbf{C. Systemic and operational threats}} \\
\midrule
Resource/\allowbreak{}Availability Exploits
& Infrastructure
& Resource Exhaustion~\cite{narajala2025securing,deng2025ai, sivaroopan2026shield}
& Infrastructure--Agent
& AV
& S, M \\
\addlinespace[2pt]
Isolation/\allowbreak{}Credential Compromise
& Infrastructure
& Sandbox Escape~\cite{guo2025systematic,anbiaee2026security}; Token Theft and Account Takeover~\cite{guo2025systematic}
& Infrastructure--Agent; Service--Agent
& AI, CF
& S \\
\addlinespace[2pt]
Component/\allowbreak{}Dependency Compromise
& Supply Chain
& Supply-Chain/Dependency Compromise ~\cite{deng2025ai,qu2026supply,jiang2026agentic}; Compromised Tools, Agents, Plugins, or Dependencies ~\cite{qu2026supply,jiang2026agentic}
& Supply Chain--Agent
& TI, AI
& S, M \\
\addlinespace[2pt]
Installation/\allowbreak{}Update Manipulation
& Supply Chain
& Installer Spoofing~\cite{guo2025systematic}; Malicious Updates and Post-Installation Modification ~\cite{guo2025systematic,bhatt2025etdi,acharya2026formal}; Tool/MCP Ecosystem Poisoning ~\cite{qu2026supply,jiang2026agentic}
& Supply Chain--Agent
& TI
& S \\
\addlinespace[2pt]
Oversight Degradation
& Governance \& Oversight
& Oversight Saturation; Monitoring Fatigue ~\cite{narajala2025securing,abdennebi2026lang}
& Human--Agent; Governance
& OA
& S, M \\
\addlinespace[2pt]
Governance Evasion
& Governance \& Oversight
& Governance Evasion and Obfuscation ~\cite{narajala2025securing}; Strategic Monitor Evasion ~\cite{greenblatt2024alignment,hubinger2024sleeper}
& Governance
& OA
& S, M \\
\bottomrule
\end{tabularx}
\par\vspace{4pt}
\raggedright
\textbf{Architecture:} S = Single-agent;
H = Hierarchical; P = Sequential/Pipeline;
C = Horizontal/Collaborative; D = Dynamic/Graph-based;
M = Multi-agent, topology not reliably classifiable; M is not an umbrella
label for H, P, C, or D.
\par\smallskip
\textbf{Security properties:} GI = Goal Integrity;
KI = State and Knowledge Integrity;
AI = Action and Authorization Integrity;
TI = Interaction and Trust Integrity;
OA = Oversight and Accountability;
CF = Confidentiality; AV = Availability;
SF = Physical Safety (for embodied actions).
\par\smallskip
\textbf{Reading the mapping:} Labels within a row need not co-occur in
one experiment. Architecture labels describe reported empirical settings,
not all theoretically applicable architectures. Omission does not imply
inapplicability. Shared labels do not establish a temporal or causal sequence.
\end{minipage}%
}
\sbox{\TaxonomyContentBox}{%
  \resizebox{\textwidth}{!}{\usebox{\TaxonomyContentBox}}%
}
\ifdim\dimexpr\ht\TaxonomyContentBox+\dp\TaxonomyContentBox\relax>0.95\textheight
  \resizebox*{!}{0.95\textheight}{\usebox{\TaxonomyContentBox}}%
\else
  \usebox{\TaxonomyContentBox}%
\fi
\endgroup
\end{table*}


\subsubsection{Memory and Knowledge Threats}
\label{sec:memory_threats}

Memory attacks are particularly important because compromise may persist
after the original adversarial interaction and influence future tasks.

\concept{Knowledge and Memory Poisoning.}
Attackers inject adversarial information into persistent memory, RAG stores,
knowledge bases, or retained experiences
\cite{narajala2025securing,zhang2025agent,he2025emerged,yu2025survey,
dong2025memory,srivastava2025memorygraft,chen2024agentpoison}. Poisoned
state may later be retrieved as trusted evidence and repeatedly influence
reasoning.

\concept{Persistent Belief Manipulation and Drift.}
Repeated retrieval and reuse of compromised information can reinforce false
beliefs or produce persistent behavioral drift across tasks or sessions
\cite{srivastava2025memorygraft}.

\concept{Shared-Memory Contamination.}
A compromised agent writes malicious state into shared memory or knowledge
stores, allowing contamination to propagate to agents that never interacted
with the original adversary
\cite{liu2026memorypropagation,torra2026memory}.

\concept{Contextual Data Leakage and Retrieval Exfiltration.}
Sensitive information contained in context, persistent memory, shared state,
or retrieval systems is disclosed directly or through tools, APIs, or other
agents
\cite{yu2025survey,wang2025unveiling,lyu2026adam,liu2025topology,
deng2025ai,qi2024follow,jiang2024rag}.

\subsubsection{Action and Tool Threats}
\label{sec:action_threats}

Action and tool threats translate compromised agent decisions into effects
on external systems.

\concept{Tool and Inventory Poisoning.}
Attackers manipulate tool descriptions, metadata, selection mechanisms, or
registered capabilities through techniques including file-based injection,
logic shadowing, tool-selection subversion, rug pulls, and malicious
registration
\cite{guo2025systematic,liu2025exploit,wang2025mcptox,
jamshidi2025securing,zhang2025mcp,bhatt2025etdi,
acharya2026formal,zhao2025mcp}.

\concept{Tool-Response Manipulation.}
A compromised tool or service returns adversarial observations that alter
subsequent reasoning, planning, or execution
\cite{deng2025ai,guo2025systematic,raza2025trism}.

\concept{Command and Execution Injection.}
Adversarially influenced arguments reach shells, interpreters, APIs, or
other execution interfaces without adequate validation, enabling
unauthorized command or code execution
\cite{yu2025survey,guo2025systematic,liu2025exploit}.

\concept{Unauthorized Action and Privilege Escalation.}
An agent is induced to perform actions beyond its intended authority
\cite{narajala2025securing,guo2025systematic,ji2026taming,
mitra2026agenticcyops}. Importantly, individually permitted operations may
compose into an unauthorized trajectory.

\concept{Multi-Tool and Cross-Service Exploitation.}
Multiple legitimate tools or services are composed to achieve a security
consequence that would not be permitted through an individual operation
\cite{guo2025systematic,dehghantanha2026sok}.

\subsection{Interaction and Trust-Boundary Threats}
\label{sec:interaction_threats}

Interaction threats describe the relationships through which adversarial
information, authority, or state enters and propagates. They complement the
functional taxonomy rather than forming mutually exclusive categories.

\subsubsection{Human--Agent Interaction}
\label{sec:human_agent_threats}

Human--agent interactions create a security boundary because instructions,
authority, sensitive information, and consequential decisions may pass
between users and autonomous agents.

\concept{Trust Manipulation.}
Human--agent trust can be exploited in both directions. Adversarial
instructions or social-engineering strategies may influence how users and
agents exchange information and authority, while a compromised or misleading
agent may exploit user trust through persuasive recommendations,
explanations, or assurances~\cite{yu2025survey}. The resulting consequences
may include disclosure of information, approval of unsafe actions, or
inappropriate delegation of authority.

\concept{Oversight and Approval Manipulation.}
Weak or bypassed human-in-the-loop controls and overly broad delegation can
allow consequential agent actions to proceed without meaningful supervision
\cite{narajala2025securing,yu2025survey}. Such failures are particularly
important when agents can invoke tools or act on external systems under
delegated human authority.

\subsubsection{Agent--Agent Interaction}
\label{sec:agent_agent_threats}

\concept{Identity Spoofing and Trust Exploitation.}
Attackers impersonate trusted agents or exploit weak identity verification
to inject instructions, observations, or delegated tasks with forged
authority
\cite{narajala2025securing,deng2025ai,he2025red}.

\concept{Inter-Agent Message Manipulation.}
Messages between legitimate agents may be intercepted, modified, replayed,
suppressed, or selectively delivered, altering the state from which
recipient agents reason~\cite{he2025red}.

\concept{Cross-Agent Threat Propagation.}
Compromised instructions, beliefs, or state propagate through messages,
delegation, intermediate outputs, or shared memory
\cite{deng2025ai,yu2025survey,nakamura2026colosseum,
liu2026memorypropagation,torra2026memory}.

\concept{Consensus Poisoning and Byzantine Behavior.}
Malicious participants manipulate voting, debate, critique, or aggregation,
or behave inconsistently across peers and execution stages. Recent work
shows that robustness to Byzantine agents depends on communication and
aggregation structure
\cite{lee2026byzantine,elmir2026byzantine}.

\concept{Collusion and Coordinated Manipulation.}
Multiple agents strategically coordinate to manipulate collective beliefs,
manufacture false consensus, conceal malicious behavior, or bypass controls
that assume independent failures
\cite{hu2026lying,zeng2026collusion}.

\subsubsection{Agent--Tool and Service Interaction}
\label{sec:agent_tool_threats}

\concept{Tool and Service Impersonation.}
Attackers may introduce malicious or impersonated tools, services, plugins,
or MCP endpoints that appear to provide trusted capabilities. Related
techniques include tool squatting, malicious tool registration, and
post-registration manipulation
\cite{guo2025systematic,bhatt2025etdi,zhao2025mcp}. Successful
impersonation may expose credentials, redirect execution, manipulate agent
state, or enable data exfiltration.

\concept{Poisoned Metadata and Capability Descriptions.}
Manipulated tool descriptions, schemas, or documentation influence tool
selection, argument construction, or planning
\cite{guo2025systematic,jamshidi2025securing,zhang2025mcp,
zhao2025mcp}.

\concept{Compromised Tool and Service Responses.}
External components return adversarial observations that the agent treats as
trusted execution results
\cite{deng2025ai,guo2025systematic,raza2025trism}. Such observations may
subsequently affect reasoning, memory, or further actions.

\concept{Credential and Delegated-Authority Abuse.}
Authentication tokens, API keys, service credentials, or delegated
capabilities are stolen or misused, allowing attackers to exercise the
agent's authority over connected services~\cite{guo2025systematic}.

\subsubsection{Agent--Environment Interaction}
\label{sec:agent_environment_threats}

\concept{Environmental Observation Manipulation.}
Attackers may modify web content, files, user-interface elements,
multimodal inputs, sensor readings, or other observations from which an
agent constructs its representation of the environment
\cite{deng2025ai,yu2025survey}.

\concept{Environmental Instruction Injection.}
Instructions embedded within observed environmental content are interpreted
as commands rather than untrusted data
\cite{greshake2023not,zhan2024injecagent,debenedetti2024agentdojo}.
Because agents autonomously retrieve and act on external information, such
content can cross the environment boundary and subsequently influence
reasoning, memory, or tool use.

\concept{Physical and Sensor Manipulation.}
Adversarial interference with cameras, microphones, sensors, or other
physical interfaces can distort an agent's perception of its environment
\cite{deng2025ai,yu2025survey}.

\concept{Unsafe Embodied Actions.}
In embodied settings, failures in task planning or environmental
interpretation can result in unsafe actions with physical consequences
\cite{yin2024safeagentbench}. This extends agent security beyond information
integrity to the safety and integrity of actions performed in the physical
environment.

\subsection{Systemic and Operational Threats}
\label{sec:systemic_threats}

Systemic threats target infrastructure, dependencies, and governance
mechanisms on which multiple agent capabilities and interactions depend.

\subsubsection{Infrastructure and Resource Threats}
\label{sec:infrastructure_threats}

\concept{Resource Exhaustion.}
Attackers consume computation, memory, context windows, API quotas, or
external-service capacity
\cite{narajala2025securing,deng2025ai,sivaroopan2026shield}.
Autonomous reasoning and tool invocation can amplify the resulting cost and
availability impact.

\concept{Sandbox Escape and Isolation Failure.}
Weaknesses in code- or tool-execution isolation are exploited to access host
files, networks, credentials, or other resources beyond the intended
boundary~\cite{guo2025systematic,anbiaee2026security}.

\concept{Credential Theft and Account Takeover.}
OAuth tokens or other credentials used by agents are stolen, allowing
attackers to inherit authority over connected external services
\cite{guo2025systematic}.

\subsubsection{Supply-Chain and Ecosystem Threats}
\label{sec:supply_chain_threats}

\concept{Dependency and Component Compromise.}
Models, libraries, APIs, tools, plugins, or other dependencies are
compromised before or during deployment
\cite{deng2025ai,qu2026supply,jiang2026agentic}.

\concept{Agent and Tool Supply-Chain Poisoning.}
Malicious agents, skills, prompts, plugins, tools, or MCP servers enter the
system through trusted distribution mechanisms. Rug-pull attacks further
allow initially benign components to become malicious after trust has been
established
\cite{guo2025systematic,bhatt2025etdi,acharya2026formal}.

\concept{Installer and Update Manipulation.}
Automated installation or update mechanisms are spoofed or compromised to
introduce malicious components or modify system configuration with
installation-time privileges~\cite{guo2025systematic}.

\subsubsection{Governance and Oversight Threats}
\label{sec:governance_threats}

\concept{Oversight Degradation.}
Excessive alerts, approval requests, actions, or execution traces can
overwhelm human or automated monitoring capacity, reducing the effectiveness
of security oversight
\cite{narajala2025securing,abdennebi2026lang}.

\concept{Governance Evasion and Obfuscation.}
Malicious behavior may be fragmented, delayed, or distributed across agents,
tools, or time so that individual events remain below detection thresholds.
Complex interactions and incomplete execution traces may further make
attribution difficult and weaken governance controls
\cite{narajala2025securing}.

\concept{Strategic Oversight Circumvention.}
Agents may alter their behavior according to monitoring, evaluation, or
approval conditions. Related work on alignment faking and sleeper behavior
illustrates the concern that evaluation-time behavior may not represent
deployment-time behavior
\cite{greenblatt2024alignment,hubinger2024sleeper}.\\

\begin{openquestion}{Answer to RQ1}
We characterize agentic AI threats through
$\mathcal{T}=\langle \mathcal{S},\mathcal{B},\mathcal{P},\mathcal{A}\rangle$, linking affected functional or system
surfaces ($\mathcal{S}$), interaction or trust boundaries ($\mathcal{B}$), violated
security properties ($\mathcal{P}$), and empirically examined architectures ($\mathcal{A}$).
This cross-dimensional representation connects complementary threat
attributes without encoding their temporal order or causal sequence.
Each dimension may contain multiple labels, allowing threats to span
several surfaces, boundaries, properties, and evaluated architectures.
\end{openquestion}
\vspace{1\baselineskip}

%% file: Section/red-teaming.tex
\section{Empirical Analysis of Agentic AI Red Teaming}
\label{sec:redteaming}

The taxonomy in Section~\ref{sec:taxonomy} characterizes the broader
security landscape of agentic AI. However, the existence of a documented
threat does not imply that it has been empirically demonstrated against an
agentic system. To address RQ2 and RQ3, we therefore separately analyze
studies that instantiate concrete attacks and evaluate their effects on
LLM-based agents.

Following the selection criteria in Section~\ref{sec:methodology}, we retain
22 empirical studies that (i) target an LLM-based agent or multi-agent
system, (ii) implement or experimentally evaluate at least one concrete
attack, (iii) provide sufficient information to characterize the target,
attacker, architecture, and experimental outcome, and (iv) release at least
one publicly accessible reproducibility artifact, such as implementation
code, attack prompts or payloads, benchmark/test data, or evaluation
scripts/configurations. This artifact criterion enables the empirical
analysis to be grounded in studies for which at least part of the
experimental methodology can be independently inspected or reused.

For each study, we map the demonstrated attacks to the
$\mathcal{T}=\langle \mathcal{S},\mathcal{B},\mathcal{P},\mathcal{A}\rangle$ representation introduced in
Section~\ref{sec:cross_mapping} and extract the demonstrated threat surface,
attacker assumptions, evaluated target architecture, experimental
environment, defense evaluation, and publicly released reproducibility
artifacts.

\subsection{Empirical Study Landscape}
\label{sec:empirical_landscape}

Table~\ref{tab:red_papers} summarizes the 22 artifact-backed empirical
red-teaming studies retained for detailed evidence analysis. Compared with
the broader threat inventory, this view distinguishes the \emph{known attack
surface} from the \emph{empirically evaluated attack surface}. For each
study, we report the demonstrated threats, attacker assumptions, target
system, evaluated architecture, experimental environment, defense
evaluation, and publicly released reproducibility artifacts. Rather than
assigning subjective low, medium, or high reproducibility ratings, we record
observable artifact availability using four categories: implementation code
(C), attack prompts or artifacts (P), released datasets or benchmarks (D),
and evaluation scripts or configurations (E).

\begin{table*}[!t]
\caption{Empirical red-teaming studies for agentic AI with publicly
released reproducibility artifacts. Architecture badges denote target-system
configurations empirically evaluated. Artifact badges denote resources
released with or maintained for each study:
C = implementation code,
P = attack prompts, payloads, adversarial inputs, or attack-generation
artifacts,
D = released dataset, benchmark, or test cases, and
E = evaluation scripts, experiment configurations, or evaluation harnesses.
\textsuperscript{\textsf{[R]}} links to the corresponding public artifact
repository or release source.}
\label{tab:red_papers}

\centering
\scriptsize
\renewcommand{\arraystretch}{1.20}
\setlength{\tabcolsep}{3pt}

\resizebox{\textwidth}{!}{%
\begin{tabular}{
    >{\raggedright\arraybackslash}p{2.0cm}
    >{\centering\arraybackslash}p{1.5cm}
    >{\raggedright\arraybackslash}p{4.0cm}
    >{\centering\arraybackslash}p{1.0cm}
    >{\raggedright\arraybackslash}p{2.7cm}
    >{\centering\arraybackslash}p{1.7cm}
    >{\raggedright\arraybackslash}p{1.8cm}
    >{\centering\arraybackslash}p{1.0cm}
    >{\centering\arraybackslash}p{2.0cm}
}
\toprule

\textbf{Study} &
\textbf{Year / Venue} &
\textbf{Demonstrated Threats} &
\textbf{Attacker} &
\textbf{Target System / Framework} &
\makecell[c]{\textbf{Empirical}\\\textbf{Arch.}} &
\textbf{Environment} &
\makecell[c]{\textbf{Defense}\\\textbf{eval.}} &
\textbf{Artifacts} \\

\midrule

Greshake et al.~\cite{greshake2023not}
&
2023 / AISec
&
Indirect prompt injection; persistent compromise;
data exfiltration; unauthorized actions
&
\atkB
&
LangChain, Bing Chat, Copilot
&
\archS
&
Web/tool
&
\defno
&
\artC\artsep\artP\artsep\artE
\hspace{1.5pt}
\repolink{https://github.com/greshake/llm-security}
\\

Zhan et al.~\cite{zhan2024injecagent}
&
2024 / ACL
&
Indirect prompt injection; direct harm; data exfiltration
&
\atkB
&
InjecAgent
&
\archS
&
Tool-integrated
&
\defno
&
\artC\artsep\artP\artsep\artD\artsep\artE
\hspace{1.5pt}
\repolink{https://github.com/uiuc-kang-lab/InjecAgent}
\\

Gu et al.~\cite{gu2024agent}
&
2024 / ICML
&
Multimodal jailbreak; cross-agent infection and propagation
&
\atkW
&
Agent Smith
&
\archD
&
Multi-agent
&
\defno
&
\artC\artsep\artP\artsep\artD\artsep\artE
\hspace{1.5pt}
\repolink{https://github.com/sail-sg/Agent-Smith}
\\

Amayuelas et al.~\cite{amayuelas2024multiagent}
&
2024 / EMNLP
&
Adversarial reasoning manipulation; trust sabotage;
collaborative error propagation
&
\atkG
&
Multi-agent debate
&
\archC
&
Multi-agent
&
\defyes
&
\artC\artsep\artP\artsep\artE
\hspace{1.5pt}
\repolink{https://github.com/amayuelas/multi-agent-attack}
\\

Debenedetti et al.~\cite{debenedetti2024agentdojo}
&
2024 / NeurIPS
&
Indirect prompt injection; malicious tool-mediated actions
&
\atkB
&
AgentDojo
&
\archS
&
Tool-integrated
&
\defyes
&
\artC\artsep\artP\artsep\artD\artsep\artE
\hspace{1.5pt}
\repolink{https://github.com/ethz-spylab/agentdojo}
\\

Chen et al.~\cite{chen2024agentpoison}
&
2024 / NeurIPS
&
Memory poisoning; knowledge-base poisoning;
persistent backdoor activation
&
\atkW
&
AgentPoison
&
\archS
&
Memory/RAG
&
\defno
&
\artC\artsep\artP\artsep\artD\artsep\artE
\hspace{1.5pt}
\repolink{https://github.com/AI-secure/AgentPoison}
\\

Yang et al.~\cite{yang2024watch}
&
2024 / NeurIPS
&
Query-, observation-, and thought-triggered agent backdoors
&
\atkW
&
AgentTuning, ToolBench
&
\archS
&
Tool-integrated
&
\defno
&
\artC\artsep\artP\artsep\artD\artsep\artE
\hspace{1.5pt}
\repolink{https://github.com/lancopku/agent-backdoor-attacks}
\\

Wang et al.~\cite{wang2024badagent}
&
2024 / ACL
&
Backdoor insertion; trigger-activated harmful actions
&
\atkW
&
BadAgent
&
\archS
&
Web/tool
&
\defno
&
\artC\artsep\artP\artsep\artE
\hspace{1.5pt}
\repolink{https://github.com/DPamK/BadAgent}
\\

Andriushchenko et al.~\cite{andriushchenko2024agentharm}
&
2025 / ICLR
&
Jailbreak; harmful autonomous tool use;
multi-step malicious behavior
&
\atkB
&
AgentHarm
&
\archS
&
Tool-integrated
&
\defyes
&
\artP\artsep\artD\artsep\artE
\hspace{1.5pt}
\repolink{https://huggingface.co/datasets/ai-safety-institute/AgentHarm}
\\

Zhang et al.~\cite{zhang2025agent}
&
2025 / ICLR
&
Direct/indirect prompt injection; memory poisoning;
Plan-of-Thought backdoor; mixed attacks
&
\atkG
&
Agent Security Bench
&
\archS
&
Tool/memory
&
\defyes
&
\artC\artsep\artP\artsep\artD\artsep\artE
\hspace{1.5pt}
\repolink{https://github.com/agiresearch/ASB}
\\

Huang et al.~\cite{huangresilience}
&
2025 / ICML
&
Faulty/malicious-agent injection; reasoning-error propagation;
cross-agent reliability degradation
&
\atkG
&
CAMEL, MetaGPT, SPP, MAD, AgentVerse
&
\archH\archsep\archP\archsep\archC
&
Multi-agent
&
\defyes
&
\artC\artsep\artP\artsep\artD\artsep\artE
\hspace{1.5pt}
\repolink{https://github.com/CUHK-ARISE/MAS-Resilience}
\\

Zhang et al.~\cite{zhang2025udora}
&
2025 / ICML
&
Dynamic reasoning hijacking; adversarial-string optimization;
indirect injection
&
\atkW
&
UDora
&
\archS
&
Web/tool
&
\defno
&
\artC\artsep\artP\artsep\artD\artsep\artE
\hspace{1.5pt}
\repolink{https://github.com/AI-secure/UDora}
\\

Fang et al.~\cite{fang2025we}
&
2025 / arXiv
&
Malicious MCP services; indirect prompt injection;
adversarial tool responses
&
\atkB
&
SafeMCP
&
\archS
&
MCP/tool
&
\defyes
&
\artC\artsep\artP\artsep\artE
\hspace{1.5pt}
\repolink{https://github.com/littlelittlenine/SafeMCP}
\\

He et al.~\cite{he2025red}
&
2025 / ACL
&
MITM communication manipulation; identity spoofing;
cross-agent message attacks
&
\atkB
&
AutoGen, CAMEL, MetaGPT, ChatDev
&
\archH\archsep\archP\archsep\archC\archsep\archD
&
Multi-agent
&
\defno
&
\artC\artsep\artP
\hspace{1.5pt}
\repolink{https://github.com/PengfeiHePower/AiTM}
\\

Gasmi et al.~\cite{gasmi2025bridging}
&
2025 / arXiv
&
Prompt injection; malicious tools; command injection;
privilege escalation; resource abuse
&
\atkB
&
Azure OpenAI, AWS Bedrock
&
\archS
&
Cloud/tool
&
\defno
&
\artC\artsep\artP\artsep\artE
\hspace{1.5pt}
\repolink{https://github.com/theconsciouslab-ai/llm-agent-security}
\\

Zou et al.~\cite{zou2025security}
&
2025 / NeurIPS
&
Prompt injection; policy violations;
adversarial agent deployment
&
\atkB
&
Agent Red Teaming (ART)
&
\archS
&
Tool-integrated
&
\defno
&
\artP\artsep\artD\artsep\artE
\hspace{1.5pt}
\repolink{https://proceedings.neurips.cc/}
\\

Atta et al.~\cite{atta2026laaf}
&
2026 / arXiv
&
Logic-layer injection; trigger execution; persistence;
evasion; trace manipulation
&
\atkB
&
LAAF
&
\archS
&
Tool/API
&
\defno
&
\artC\artsep\artP\artsep\artE
\hspace{1.5pt}
\repolink{https://github.com/qorvexconsulting1/laaf-V2.0}
\\

Kavathekar et al.~\cite{kavathekar2026tamas}
&
2026 / ACL
&
Multi-agent adversarial attacks; trust/coordination manipulation;
cross-agent failures
&
\atkB
&
TAMAS
&
\archH\archsep\archP\archsep\archC
&
Multi-agent/tool
&
\defno
&
\artC\artsep\artP\artsep\artD\artsep\artE
\hspace{1.5pt}
\repolink{https://github.com/microsoft/TAMAS}
\\

Jiang et al.~\cite{jiang2026agentlab}
&
2026 / arXiv
&
Intent hijacking; tool chaining; task injection;
objective drifting; memory poisoning
&
\atkB
&
AgentLAB
&
\archS
&
Long-horizon/tool
&
\defyes
&
\artC\artsep\artP\artsep\artD\artsep\artE
\hspace{1.5pt}
\repolink{https://github.com/TanqiuJiang/AgentLAB}
\\

Zhang et al.~\cite{zhang2025mcp}
&
2026 / ICLR
&
MCP-specific attacks across planning,
tool invocation, and response handling
&
\atkB/\atkG
&
MCP Security Bench
&
\archS
&
MCP/tool
&
\defno
&
\artC\artsep\artP\artsep\artD\artsep\artE
\hspace{1.5pt}
\repolink{https://github.com/dongsenzhang/MSB}
\\

Wang et al.~\cite{wang2025mcptox}
&
2026 / AAAI
&
Tool-description poisoning; instruction hijacking;
unauthorized tool actions
&
\atkB
&
MCPTox
&
\archS
&
MCP/tool
&
\defno
&
\artC\artsep\artP\artsep\artD\artsep\artE
\hspace{1.5pt}
\repolink{https://github.com/zhiqiangwang4/MCPTox-Benchmark}
\\

Liu et al.~\cite{liu2025topology}
&
2026 / ACL
&
Cross-agent memory leakage; privacy extraction;
topology-conditioned propagation
&
\atkB
&
MAMA
&
\archH\archsep\archP\archsep\archC\archsep\archD
&
Multi-agent/memory
&
\defno
&
\artC\artsep\artP\artsep\artD\artsep\artE
\hspace{1.5pt}
\repolink{https://github.com/llll121/mama-eval}
\\

\bottomrule
\end{tabular}%
}

\vspace{4pt}

\raggedright
\scriptsize

\textbf{Attacker knowledge:}\quad
\atkB~Black-box;
\hspace{4pt}
\atkG~Gray-box;
\hspace{4pt}
\atkW~White-box.

\vspace{2pt}

\textbf{Architecture evidence:}\quad
\archS~Single-agent
\hspace{4pt}
\archH~Hierarchical
\hspace{4pt}
\archP~Sequential/Pipeline
\hspace{4pt}
\archC~Horizontal/Collaborative
\hspace{4pt}
\archD~Dynamic/Graph-based
\hspace{4pt}
\archM~Multi-agent, topology not reliably classifiable.

\vspace{2pt}

\textbf{Defense eval.:}\quad
\defyes~At least one concrete defense experimentally evaluated;
\hspace{4pt}
\defno~No concrete defense evaluation.

\vspace{2pt}

\textbf{Reproducibility artifacts:}\quad
\artC~Code
\hspace{4pt}
\artP~Prompts/attack artifacts
\hspace{4pt}
\artD~Dataset/benchmark
\hspace{4pt}
\artE~Evaluation scripts/configurations;
\quad
\textsuperscript{\textsf{[R]}}~Public artifact repository/source.

\end{table*}

\subsection{Threat and Architecture Coverage}
\label{sec:redteam_coverage}

Within the 22 artifact-backed studies reviewed, coverage concentrates
on prompt injection, jailbreaks, memory poisoning, and tool-mediated
attacks. Human--Agent oversight, systemic governance, and cross-session
compromise are less represented in this subset. This distribution
characterizes the selected evidence rather than the full extent of
empirical research, and greater representation should not itself be
interpreted as greater evaluation maturity.

Table~\ref{tab:empirical_mapping_examples} applies the proposed
framework to selected evaluations from three studies in the corpus.
The attack descriptions and configurations follow the original papers;
the surface, boundary, and property labels are our classifications
under Section~\ref{sec:cross_mapping}. Each row specifies an evaluated
setting rather than aggregating all architectures examined by a study.
The mappings associate security-relevant attributes without encoding
temporal order or causal sequences.

\begin{table*}[t]
\centering
\caption{Cross-dimensional mapping of selected empirical evaluations.
Each row identifies a reported attack, its classification under the
proposed framework, and a configuration evaluated in the cited study.
The rows do not exhaust each study's attacks or architecture coverage.}
\label{tab:empirical_mapping_examples}
\small
\setlength{\tabcolsep}{4pt}
\renewcommand{\arraystretch}{1.15}
\begin{tabularx}{\textwidth}{
    @{}
    >{\raggedright\arraybackslash}p{0.12\textwidth}
    >{\raggedright\arraybackslash}X
    >{\raggedright\arraybackslash}p{0.15\textwidth}
    >{\raggedright\arraybackslash}p{0.12\textwidth}
    >{\raggedright\arraybackslash}p{0.18\textwidth}
    >{\raggedright\arraybackslash}p{0.15\textwidth}
    @{}
}
\toprule
\textbf{Study} &
\textbf{Reported attack} &
\textbf{Surface ($\mathcal{S}$)} &
\textbf{Boundary ($\mathcal{B}$)} &
\textbf{Property ($\mathcal{P}$)} &
\textbf{Evaluated setting ($\mathcal{A}$)} \\
\midrule

AgentDojo~\cite{debenedetti2024agentdojo}
&
Instructions injected into tool-returned data redirect the agent
toward an attacker-defined task.
&
Reasoning and Planning; Action and Tools
&
Agent--Tool
&
Goal Integrity; Action and Authorization Integrity
&
Single-agent tool-calling configuration (S)
\\

\addlinespace
AiTM~\cite{he2025red}
&
Intercepted inter-agent messages induce attacker-specified
transformations of MMLU answer labels.
&
Reasoning and Planning
&
Agent--Agent
&
Goal Integrity; Interaction and Trust Integrity
&
Three-agent chain in AutoGen and CAMEL (P)
\\

\addlinespace
MAMA~\cite{liu2025topology}
&
An information-seeking agent extracts synthetic PII initially
available only to a target agent.
&
Memory and Knowledge
&
Agent--Agent
&
Confidentiality
&
Complete communication graph with 4--6 agents (C)
\\

\bottomrule
\end{tabularx}

\smallskip
\begin{minipage}{\textwidth}
\footnotesize
\textbf{Architecture labels:}
S = Single-agent;
P = Sequential/Pipeline;
C = Horizontal/Collaborative.
The Agent--Tool boundary denotes the interface through which
tool-returned content enters the agent, irrespective of the
underlying document or service that supplies that content.
\end{minipage}
\end{table*}

The mapping separates shared attack surfaces from distinct security
consequences. AgentDojo and AiTM both involve reasoning, but adversarial
instructions enter through tool outputs and inter-agent messages,
respectively. AiTM and MAMA share an Agent--Agent boundary, yet the
selected evaluations concern behavioral redirection and confidentiality.
A boundary-level classification alone would obscure this difference.

The corresponding outcome measures also differ. AgentDojo uses
task-specific checks to determine whether attacker objectives are
achieved in its execution environments. AiTM's selected MMLU attack
is assessed through the prescribed answer-label transformation,
rather than an executed external action. MAMA measures recovery of
synthetic PII from attacker outputs using exact matching supplemented
by an inference-based judge
\cite{debenedetti2024agentdojo,he2025red,liu2025topology}.
These endpoints should remain distinct when comparing empirical
coverage and interpreting attack success.

Architecture coverage is uneven: Table~\ref{tab:red_papers} records
16 single-agent studies and six multi-agent studies. Within the latter,
AiTM compares Chain, Tree, Complete, and Random communication
structures, while MAMA evaluates complete, circle, chain, tree,
star-pure, and star-ring topologies
\cite{he2025red,liu2025topology}.
Their results also caution against assigning a universal security
ranking to topologies. Chains are particularly vulnerable to AiTM's
communication attacks, whereas chain and tree configurations are
generally among the lower-leakage settings in MAMA.
This contrast does not establish contradictory findings: the studies
differ in attacker control, communication protocols, and measured
security properties. Architecture comparisons must therefore retain
these experimental conditions.

\emph{Temporal scope} requires similar precision. A multi-step
execution, propagation between agents, and persistence across sessions
are different evaluation conditions. AgentLAB examines intent
hijacking, tool chaining, task injection, objective drifting, and
memory poisoning over extended interactions. Its memory-poisoning
procedure separates the injection of malicious memories from their
retrieval and exploitation in a subsequent target scenario
\cite{jiang2026agentlab}.
This distinction makes the relationship between retained state and
later behavior explicit. Evaluations should accordingly report the
interaction horizon, the state retained between phases, and whether
compromise survives a task or session boundary. Immediate attack
success alone does not establish persistent system compromise.

\subsection{Benchmarks and Evaluation Practices}
\label{sec:benchmark_analysis}

A central challenge in comparing agent-security research is the diversity of
evaluation environments. Some studies construct attack-specific settings,
whereas others introduce reusable benchmarks that standardize tasks, tools,
adversarial conditions, and success criteria. Table~\ref{tab:benchmarks} summarizes \textit{11 representative security benchmarks} spanning indirect prompt injection, memory compromise, harmful
autonomous behavior, embodied safety, multi-agent security, long-horizon
attacks, and MCP/tool-security evaluation.
The benchmarks differ substantially in evaluation scale, target architecture,
environmental realism, and outcome definitions; consequently, their reported
metrics should not be interpreted as directly comparable.

\begin{table*}[!t]
\caption{Representative benchmarks and evaluation environments for agentic
AI security. Architecture badges denote the target-system configurations
empirically evaluated. Evaluation scale is reported using the native unit of
each benchmark (e.g., tasks, security cases, or attack instances).
\textsuperscript{\textsf{[R]}} links to the corresponding public benchmark
repository or release source.}
\label{tab:benchmarks}

\centering
\scriptsize
\renewcommand{\arraystretch}{1.20}
\setlength{\tabcolsep}{2.7pt}

\resizebox{\textwidth}{!}{%
\begin{tabular}{
    >{\raggedright\arraybackslash}p{2cm}
    >{\centering\arraybackslash}p{1cm}
    >{\raggedright\arraybackslash}p{3cm}
    >{\raggedright\arraybackslash}p{2.0cm}
    >{\centering\arraybackslash}p{1.0cm}
    >{\raggedright\arraybackslash}p{2.1cm}
    >{\raggedright\arraybackslash}p{3.2cm}
    >{\raggedright\arraybackslash}p{2.2cm}
}
\toprule

\textbf{Benchmark} &
\textbf{Year / Venue} &
\textbf{Primary Security Focus} &
\makecell[c]{\textbf{Evaluation}\\\textbf{Scale}} &
\makecell[c]{\textbf{Empirical}\\\textbf{Arch.}} &
\textbf{Target Setting} &
\textbf{Evaluation Environment} &
\textbf{Typical Metrics} \\

\midrule

InjecAgent~\cite{zhan2024injecagent}
\hspace{1pt}
\repolink{https://github.com/uiuc-kang-lab/InjecAgent}
&
2024 / ACL
&
Indirect prompt injection; direct harm; data exfiltration
&
1,054 security cases
&
\archS
&
Tool-integrated agents
&
17 user tools; 62 attacker tools; 30 LLM agents
&
ASR; direct harm; exfiltration
\\

AgentDojo~\cite{debenedetti2024agentdojo}
\hspace{1pt}
\repolink{https://github.com/ethz-spylab/agentdojo}
&
2024 / NeurIPS
&
Indirect prompt injection; attack/defense evaluation
&
97 tasks; 629 security cases
&
\archS
&
Tool-using agents
&
Dynamic email, banking, travel, and productivity environments
&
Utility; ASR; security violation
\\

AgentPoison~\cite{chen2024agentpoison}
\hspace{1pt}
\repolink{https://github.com/AI-secure/AgentPoison}
&
2024 / NeurIPS
&
Memory and knowledge-base poisoning; persistent backdoors
&
Attack-specific evaluations
&
\archS
&
Memory/RAG agents
&
Long-term memory and knowledge-retrieval settings
&
ASR; benign utility; retrieval effectiveness
\\

SafeAgentBench~\cite{yin2024safeagentbench}
\hspace{1pt}
\repolink{https://github.com/safeagentbench/SafeAgentBench}
&
2024 / arXiv
&
Embodied-agent safety; hazardous task planning
&
750 tasks
&
\archS
&
Embodied agents
&
10 hazard categories; 3 task types; simulated embodied environment
&
Task success; rejection; safety
\\

AgentHarm~\cite{andriushchenko2024agentharm}
\hspace{1pt}
\repolink{https://huggingface.co/datasets/ai-safety-institute/AgentHarm}
&
2025 / ICLR
&
Jailbreak; harmful autonomous tool use
&
110 base tasks; 440 augmented
&
\archS
&
Tool-using agents
&
11 harm categories; multi-step malicious tasks
&
Refusal; harmful-task completion
\\

Agent Security Bench (ASB)~\cite{zhang2025agent}
\hspace{1pt}
\repolink{https://github.com/agiresearch/ASB}
&
2025 / ICLR
&
Prompt injection; memory poisoning; planning backdoors;
attack/defense evaluation
&
50 benign; 400 attack tasks
&
\archS
&
Tool/memory agents
&
10 scenarios; 10 agents; 400+ tools;
27 attack/defense methods
&
ASR; utility; security--utility trade-off
\\

AI Agent Red-Team Benchmark~\cite{zou2025security}
\hspace{1pt}
\repolink{https://proceedings.neurips.cc/paper_files/paper/2025/hash/73368bc7644c054b5bcc6490a8f2fb1c-Abstract-Datasets_and_Benchmarks_Track.html}
&
2025 / NeurIPS
&
Large-scale adversarial robustness; prompt injection;
policy violations
&
1.8M attacks
&
\archS
&
Frontier LLM agents
&
22 frontier LLMs; 44 realistic deployment scenarios
&
ASR; transferability; policy violation
\\

TAMAS~\cite{kavathekar2026tamas}
\hspace{1pt}
\repolink{https://github.com/microsoft/TAMAS}
&
2026 / ACL
&
Adversarial multi-agent security; coordination and trust attacks
&
300 adversarial; 100 harmless tasks
&
\archH\archsep\archP\archsep\archC
&
Multi-agent systems
&
5 scenarios; 6 attack types; 211 tools;
10 LLMs; 3 interaction configurations
&
ERS; robustness; task effectiveness
\\

AgentLAB~\cite{jiang2026agentlab}
\hspace{1pt}
\repolink{https://github.com/TanqiuJiang/AgentLAB}
&
2026 / arXiv
&
Adaptive long-horizon attacks; persistent agent compromise
&
644 attack cases
&
\archS
&
Long-horizon agents
&
28 environments; 5 attack families;
multi-turn agent--environment trajectories
&
ASR; task outcome; defense effectiveness
\\

MCP Security Bench (MSB)~\cite{zhang2025mcp}
\hspace{1pt}
\repolink{https://github.com/dongsenzhang/MSB}
&
2026 / ICLR
&
MCP-specific attacks across planning, tool invocation,
and response handling
&
2,000 attack instances
&
\archS
&
MCP/tool agents
&
10 domains; 405 tools; 12 attack categories;
9 LLM agents
&
ASR; NRP; task performance
\\

MCPTox~\cite{wang2025mcptox}
\hspace{1pt}
\repolink{https://github.com/zhiqiangwang4/MCPTox-Benchmark}
&
2026 / AAAI
&
Tool poisoning; malicious MCP metadata and instructions
&
1,348 malicious cases
&
\archS
&
MCP/tool agents
&
45 live MCP servers; 353 authentic tools;
10 risk categories
&
ASR; refusal rate; tool-poisoning robustness
\\

\bottomrule
\end{tabular}%
}

\vspace{4pt}

\raggedright
\scriptsize
\textbf{Architecture evidence:}\quad
\archS~Single-agent
\hspace{4pt}
\archH~Hierarchical
\hspace{4pt}
\archP~Sequential/Pipeline
\hspace{4pt}
\archC~Horizontal/Collaborative
\hspace{4pt}
\archD~Dynamic/Graph-based.
\quad
\textsuperscript{\textsf{[R]}}~Public benchmark repository/source.

\end{table*}

The benchmark landscape shows a clear expansion in both attack-surface
coverage and evaluation realism. InjecAgent provides a focused benchmark for
indirect prompt injection across heterogeneous tool interactions
\cite{zhan2024injecagent}, while AgentDojo introduces dynamic environments
in which attack success can be evaluated jointly with benign task utility
\cite{debenedetti2024agentdojo}. AgentPoison isolates persistent compromise
of memory and knowledge retrieval~\cite{chen2024agentpoison}, whereas
SafeAgentBench extends evaluation to embodied settings in which unsafe
planning may result in hazardous physical actions
\cite{yin2024safeagentbench}.

AgentHarm shifts the evaluation target from harmful text generation toward
whether autonomous agents can successfully complete malicious multi-step
tasks~\cite{andriushchenko2024agentharm}. ASB broadens coverage across
system prompts, user prompts, tool use, memory retrieval, planning, and
defense evaluation~\cite{zhang2025agent}, while large-scale public red
teaming evaluates adversarial robustness across diverse frontier models and
agent configurations~\cite{zou2025security}.

More recent benchmarks increasingly target system properties that are
specific to agentic execution. TAMAS evaluates adversarial behavior across
multiple multi-agent coordination structures
\cite{kavathekar2026tamas}, while AgentLAB targets adaptive attacks over
longer agent--environment trajectories~\cite{jiang2026agentlab}. MCP Security Bench evaluates attacks across the MCP tool-use pipeline,
including planning, tool invocation, and response handling
~\cite{zhang2025mcp},
whereas MCPTox focuses specifically on tool poisoning against real-world MCP
servers and authentic tool ecosystems~\cite{wang2025mcptox}. Together,
these benchmarks extend evaluation beyond model-level prompt robustness
toward coordination, persistence, tool trust, and system-level effects.

Despite this progress, benchmark coverage remains fragmented. Existing
suites specialize in different threat families, architectures, environments,
and outcome definitions, making results difficult to compare directly.
Attack Success Rate (ASR) remains common, but the same term may represent
generation of prohibited text, selection of a malicious tool, disclosure of
information, completion of a harmful task, manipulation of another agent, or
modification of an external system state.

For agentic systems, evaluation should therefore distinguish at least three
levels of attack success:
\emph{model compromise}, where adversarial content changes the model
response;
\emph{agent compromise}, where the attack changes the agent's goal, state,
plan, or selected action; and
\emph{system impact}, where the attack produces a verifiable unauthorized,
harmful, or persistent state change.
This distinction avoids treating textual compliance and executed harm as
equivalent security outcomes.

\subsection{Defense Evaluation}
\label{sec:defense_analysis}

Seven of the 22 reviewed studies report a concrete defense evaluation.
This count establishes the presence of defense experiments, not their
coverage, adaptivity, or effectiveness against different attacks.
Evaluations should distinguish protection against a tested attack
implementation from evidence supporting a broader security claim.

AgentDojo and ASB support comparisons of attacks and defenses within
common tasks, including their effects on benign task
performance~\cite{debenedetti2024agentdojo,zhang2025agent}.
Such comparisons should report both \emph{security} and \emph{utility},
alongside operational costs such as model calls, latency, and human
intervention. Restricting tools or rejecting external information may
reduce attack success while also impairing legitimate tasks.

The coverage limitations discussed in
Section~\ref{sec:redteam_coverage} motivate defense evaluation for
persistent compromise, multi-agent interactions, and systemic threats.
Tests should specify attacker knowledge, access, and budgets, and
whether attackers adapt to the evaluated defense.

Defense evaluation should also distinguish the stages at which controls
operate. Input filtering, provenance checks, authorization enforcement,
memory-integrity checks, and trajectory monitoring address different
points in agent execution. Their effectiveness should be assessed across
action sequences, since individually permitted actions may collectively
violate security constraints.\\

\begin{openquestion}{Answer to RQ2}
Within the 22 artifact-backed studies reviewed, coverage concentrates
on prompt/reasoning, memory, and tool-mediated attacks; 16 studies
evaluate single-agent systems. Cross-session compromise, Human--Agent
oversight, and systemic threats receive less coverage in this subset.
These findings characterize the selected corpus rather than establish
field-wide under-evaluation.
\end{openquestion}
\vspace{1\baselineskip}

\subsection{Reproducibility and Evidence Maturity}
\label{sec:evidence_maturity}

The 22-study empirical subset includes only studies with at least one
public artifact. Table~\ref{tab:red_papers} distinguishes implementation
code (C), attack prompts, payloads, or generation artifacts (P), datasets,
benchmarks, or test cases (D), and evaluation scripts, configurations,
or harnesses (E). This selection excludes studies without public
artifacts, even when their publications document empirical findings.
Coverage gaps therefore concern the selected subset, while our artifact
analysis neither estimates field-wide availability nor establishes
successful independent reproduction.

Artifact availability supports inspection but does not guarantee
reproducibility. Releases range from selected prompts or code to complete
benchmark environments. Reproduction may still depend on proprietary
models, changing APIs, external services, and nondeterministic execution.
Artifact types should therefore be distinguished from the completeness
and executability of the released materials.

Agentic systems also require capturing \emph{execution state}.
Identical prompts and models can yield different trajectories when
memory, tool responses, or agent interactions change. Reproducibility
therefore requires documenting model versions and settings, prompts,
tool versions, permissions, initial memory and environment state,
agent topology, and execution traces.

Outcome definitions further complicate comparison. The reviewed
benchmarks measure refusal, textual compliance, tool selection, task
completion, benign utility, and external state changes.
Evaluation should distinguish \emph{model influence},
\emph{agent redirection}, and \emph{verified system impact},
rather than treating these outcomes as equivalent attack successes.

These observations highlight the need to complement threat discovery
with evaluation across diverse architectures, extended interactions,
adaptive defenses, and reproducible environments.\\

\begin{openquestion}{Answer to RQ3}
Evaluation maturity varies across the reviewed studies and benchmarks.
Heterogeneous metrics, limited adaptive defense evaluation,
short-horizon experiments, architectural imbalance, and incomplete
execution-state capture constrain comparison and reproducibility.
Public artifacts support inspection, but their availability alone
does not establish reproducible results.
\end{openquestion}
\vspace{1\baselineskip}

%% file: Section/discussion.tex
\section{Research Gaps and Future Directions}
\label{sec:discussion}

The answers to RQ1--RQ3 identify uneven coverage within the reviewed
evidence. The selected studies emphasize prompt manipulation,
jailbreaks, memory poisoning, and tool-mediated attacks, with less
coverage of Human--Agent oversight, systemic dependencies, and
cross-session compromise. These observations motivate 13 open
research questions addressing evaluation coverage and methodology;
they do not establish the absence of empirical research beyond
the selected corpus.

\subsection{Evaluation Maturity}
\label{sec:discussion_maturity}

The selected corpus includes 16 single-agent and six multi-agent
studies; seven report defense experiments. Ten of the 11 benchmarks
target single-agent configurations. These counts describe coverage,
not defense adaptivity, experimental realism, or reproducibility.

\concept{Evidence quality.}
Evaluation maturity should reflect model and architecture coverage,
environmental realism, attacker assumptions, temporal scope, outcome
verification, defense evaluation, and reproducibility support.
These attributes require explicit assessment beyond publication
counts or artifact availability.\\

\begin{openquestion}{Open Question OQ1}
How can evidence maturity be assessed consistently across threats
with different attacker assumptions, architectures, and evaluation
settings?
\end{openquestion}
\vspace{1\baselineskip}

\concept{Verified security outcomes.}
A manipulated response does not establish agent redirection or an
executed unauthorized action. Following
Section~\ref{sec:benchmark_analysis}, evaluations should distinguish
model compromise, agent compromise, and verified system impact,
using operational criteria appropriate to each claimed consequence.\\

\begin{openquestion}{Open Question OQ2}
How can measures of model compromise, agent compromise, and system
impact be defined and validated for comparison across tasks and
environments?
\end{openquestion}
\vspace{1\baselineskip}

\concept{Architecture-aware evaluation.}
The multi-agent studies include comparisons of coordination
structures, but findings from one configuration do not establish
security in another. Comparisons should specify topology, delegation,
shared state, communication protocols, and attacker placement,
distinguishing graph-based organization from runtime topology changes.\\

\begin{openquestion}{Open Question OQ3}
Which security findings generalize across agent architectures,
and which depend on topology, delegation, shared state, or
coordination mechanisms?
\end{openquestion}
\vspace{1\baselineskip}

\subsection{Runtime Security, Misalignment, and Persistent Compromise}
\label{sec:discussion_runtime}

The empirical literature remains strongly concentrated on attacks that
produce relatively immediate effects. Increasingly autonomous and long-lived
agents introduce a different security problem in which compromise may emerge
gradually, persist through state, or become visible only after a sequence of
individually benign actions.

\concept{Runtime misalignment and trajectory-level security.}
Current safeguards predominantly assess prompts, outputs, or individual tool
calls. Such mechanisms are less suited to detecting gradual objective drift,
reward or specification gaming, deceptive behavior, or trajectories in
which individually permissible actions collectively violate the intended
goal. Runtime security therefore requires reasoning over the agent's goal,
intermediate state, execution history, delegated authority, and downstream
consequences rather than evaluating actions independently.\\

\begin{openquestion}{Open Questions OQ4--OQ5}
\textbf{OQ4:} How can runtime monitors distinguish legitimate goal
adaptation from adversarial goal drift or strategic misalignment without
substantially restricting useful agent autonomy?

\smallskip

\textbf{OQ5:} Can trajectory-level security invariants detect harmful
multi-step behavior before irreversible actions occur when each individual
action appears benign?
\end{openquestion}
\vspace{1\baselineskip}

\concept{Persistent and cross-session compromise.}
Memory poisoning changes the security model because adversarial influence
may survive the interaction through which it was introduced. Long-lived
agents therefore require mechanisms for memory provenance, integrity
verification, rollback, expiry, compartmentalization, and selective
forgetting. The problem becomes more difficult when agents autonomously
summarize, consolidate, or rewrite memories over time.\\

\begin{openquestion}{Open Question OQ6}
How can legitimate learning and memory adaptation be distinguished from
adversarially induced belief drift, and how can compromised state be safely
rolled back without destroying useful learned information?
\end{openquestion}
\vspace{1\baselineskip}

\concept{Deceptive and trigger-dependent behavior.}
Sleeper and strategically deceptive behavior challenge evaluation
methodologies that assume test-time behavior represents deployment-time
behavior. Hidden triggers, delayed activation, monitor-aware behavior, and
long interaction histories may cause failures to remain invisible during
conventional testing
\cite{hubinger2024sleeper,greenblatt2024alignment,
betley2025emergent}. This motivates evaluation not only of whether agents
behave safely under observation, but also whether their behavior changes
when monitoring or deployment conditions differ.\\

\begin{openquestion}{Open Question OQ7}
How can security evaluations reliably detect deceptive or trigger-dependent
agent behavior when activation depends on hidden triggers, long interaction
histories, or the agent's awareness of being monitored?
\end{openquestion}
\vspace{1\baselineskip}

\subsection{Multi-Agent and Human--Agent Security}
\label{sec:discussion_multiagent}

Multi-agent systems create security dependencies that cannot be captured by
replicating single-agent tests across several models. Their security depends
on identity, delegation, communication, shared state, coordination
mechanisms, and assumptions about other participants. At the same time,
humans remain integral to agentic systems as principals, supervisors,
authorization authorities, and potential targets.

\concept{Emergent multi-agent attacks.}
Byzantine behavior, consensus poisoning, collusion, and cross-agent
propagation remain comparatively underexplored
\cite{lee2026byzantine,elmir2026byzantine,hu2026lying,
kavathekar2026tamas}. Future work should evaluate how compromise depends on
communication topology, the number and placement of malicious agents,
delegation structure, aggregation mechanisms, and persistence of shared
state. Security properties observed in collaborative peer systems should
not be assumed to generalize to hierarchical, pipeline, or dynamic
architectures.\\

\begin{openquestion}{Open Question OQ8}
How do the number, position, and coordination strategy of compromised agents
affect attack propagation and collective decisions across hierarchical,
pipeline, collaborative, and dynamic architectures?
\end{openquestion}
\vspace{1\baselineskip}

\concept{Shared-state integrity and provenance.}
Collaborative memories and common knowledge stores improve coordination but
also create persistent propagation surfaces. Recent work illustrates how
malicious state can influence agents that were not directly exposed to the
original adversary
\cite{liu2026memorypropagation,torra2026memory}. Important directions
include provenance-aware memory, constrained write authority, cross-agent
validation, contamination detection, and recovery from poisoned shared
state.\\

\begin{openquestion}{Open Question OQ9}
How can cross-agent provenance identify the origin, propagation path, and
downstream influence of poisoned messages, beliefs, or shared-memory
entries?
\end{openquestion}
\vspace{1\baselineskip}

\concept{Identity, delegation, and authority.}
Multi-agent systems require stronger notions of identity and authorization
than simple message exchange. Authenticated agent identities, scoped
delegation, privilege attenuation, revocation, and provenance across
delegation chains become particularly important in hierarchical systems,
where compromise of a coordinator may implicitly transfer malicious
authority to many downstream agents.

\concept{Human--Agent trust and meaningful oversight.}
Human oversight is frequently treated as a defense, yet humans are also part
of the attack surface. Agents may exploit automation bias, generate
misleading explanations, overwhelm approval mechanisms, or encourage users
to delegate excessive authority. Conversely, attackers may socially
engineer users into authorizing malicious agent actions. Simply inserting an
approval dialog therefore does not establish meaningful human control.

\begin{openquestion}{Open Question OQ10}
When does human oversight meaningfully reduce agentic risk, and when do
automation bias, approval fatigue, inadequate context, or unsafe delegation
make the human another exploitable component of the system?
\end{openquestion}
\vspace{1\baselineskip}

\subsection{System-Level, Supply-Chain, and Cyber--Physical Security}
\label{sec:discussion_system}

As agents shift from language generation to autonomous execution, their security increasingly depends on software ecosystems, orchestration infrastructure, external services, and physical environments. These surfaces receive less coverage than prompt- and model-level attacks in the reviewed evidence.

\concept{Agentic supply chains and continuous component trust.}
Modern agents increasingly depend on third-party tools, plugins, MCP servers,
models, reusable skills, prompts, libraries, and external services. Existing
work demonstrates several forms of malicious tool behavior, but systematic
evaluation of provenance, update integrity, dependency compromise, and
post-installation behavior remains limited. Moreover, a component should not
necessarily remain trusted simply because it was benign when first
installed: rug-pull and update attacks show that behavior may change after
trust has been established.\\

\begin{openquestion}{Open Question OQ11}
How can dynamically discovered agents, tools, skills, plugins, and MCP
servers be continuously authenticated, monitored, and revoked when their
behavior or implementation changes after deployment?
\end{openquestion}
\vspace{1\baselineskip}

\concept{Infrastructure and orchestration security.}
Multi-agent platforms increasingly rely on orchestrators, message buses,
credential stores, memory services, sandboxes, and cloud execution
environments. Compromise of these components may affect multiple agents
simultaneously. Red teaming should therefore target the orchestration and
execution plane in addition to individual agents, including message routing,
task assignment, shared state, credential management, and isolation
boundaries.\\

\concept{Cyber--physical agent systems.}
Agents connected to robotics, vehicles, laboratory automation, industrial
control, or other physical systems create consequences that cannot be
reversed as easily as generated text. Evaluation in these settings should
consider adversarial observations, sensor manipulation, unsafe action
sequences, delayed feedback, and interactions between digital compromise and
physical state. Security metrics must similarly move beyond textual attack
success toward physical safety and verifiable environmental consequences.\\

\begin{openquestion}{Open Question OQ12}
What security invariants, runtime safeguards, and recovery mechanisms are
required when autonomous agent actions modify irreversible or
safety-critical physical state?
\end{openquestion}
\vspace{1\baselineskip}

\concept{Safe recovery and containment.}
Increasing autonomy also raises the question of what happens after compromise
is detected. Important mechanisms include credential revocation, memory
rollback, safe-state restoration, interruption of delegated tasks, isolation
of compromised agents, and recovery of shared state. Recovery is especially
challenging where malicious actions have already propagated across agents or
modified external systems.\\

\subsection{Toward Reproducible and Standardized Evaluation}
\label{sec:discussion_reproducibility}

The benchmarks reviewed in Section~\ref{sec:benchmark_analysis} vary
in threats, architectures, environments, and outcome definitions,
limiting direct comparison across attacks and defenses.

\concept{Standardized threat and attacker models.}
The proposed representation,
$\mathcal{T}=\langle\mathcal{S},\mathcal{B},\mathcal{P},\mathcal{A}\rangle$,
links affected surfaces, interaction boundaries, violated security
properties, and reported empirical architectures. Attacker knowledge,
capabilities, privileges, temporal scope, and operational success
criteria are separate experimental attributes that benchmarks should
report alongside this mapping.

\concept{Comparable security metrics.}
Attack Success Rate requires an explicit definition of success.
Evaluations should distinguish model influence, agent redirection,
and verified system impact, and report benign task utility alongside
attack effectiveness. Where relevant, they should also measure
persistence, propagation, defense overhead, false-positive rates,
and the severity of resulting state changes.

\concept{Reproducibility and environment state.}
Reproducibility requires documenting both the agent configuration
and its execution environment. Releases should include model versions
and settings, prompts, tool schemas and versions, permissions,
initial memory, environment snapshots, agent topology, execution
traces, and evaluation scripts where possible. Snapshotting and
replay can support controlled comparisons, while dependencies on
changing services and nondeterministic execution should be documented.\\

\begin{openquestion}{Open Question OQ13}
How can agent-security benchmarks preserve reproducibility and
historical comparability as models, tools, environments, and
multi-agent interactions change?
\end{openquestion}
\vspace{1\baselineskip}

\concept{Continuous benchmarks and adaptive defenses.}
Maintained evaluation suites should preserve versioned tasks,
configurations, and results while testing updated systems.
Defense evaluation should include adaptive attackers that respond
to the deployed mitigation, with attacker access and budgets
explicitly reported.\\

These directions connect threat modeling to measurable security
outcomes. The research agenda emphasizes evaluation across
architectures, interaction horizons, and system boundaries,
supported by explicit threat models, comparable metrics,
and reproducible execution records.

%% file: Section/conclusion.tex
\section{Conclusion}
\label{sec:conclusion}

Agentic AI extends LLM security to persistent state, autonomous actions,
delegated authority, and interactions with other agents and external
environments. Through a structured review of 66 studies, we developed
a cross-dimensional framework linking affected system surfaces, trust
boundaries, security properties, and empirically examined architectures.
Our analysis of 22 artifact-backed red-teaming studies and 11 representative
security benchmarks characterizes the coverage and maturity of the
reviewed evidence. Within this subset, coverage concentrates on
prompt/reasoning, memory, and tool-mediated attacks, while persistent,
complex multi-agent, Human--Agent, systemic, and long-horizon threats
receive less attention. Variation in architectures, metrics, defense
evaluation, and artifact completeness limits comparison and
reproducibility. These findings concern the selected studies rather
than the full extent of empirical research.
The 13 open research questions provide an agenda for complementing
threat discovery with more systematic, realistic, and reproducible
security evaluation.

%% file: References.bib
@article{deng2025ai,
  author  = {Deng, Zehang and Guo, Yongjian and Han, Changzhou and Ma, Wanlun
             and Xiong, Junwu and Wen, Sheng and Xiang, Yang},
  title   = {{AI} Agents Under Threat: A Survey of Key Security Challenges
             and Future Pathways},
  journal = {ACM Computing Surveys},
  volume  = {57},
  number  = {7},
  year    = {2025},
  pages   = {1--36},
  doi     = {10.1145/3716628}
}

@article{shahriar2025agentic,
  author  = {Shahriar, Asif and Rahman, Md Nafiu and Ahmed, Sadif
             and Sadeque, Farig and Parvez, Md Rizwan},
  title   = {A Survey on Agentic Security: Applications, Threats and Defenses},
  journal = {arXiv preprint arXiv:2510.06445},
  year    = {2025},
  doi     = {10.48550/arXiv.2510.06445}
}

@article{chhabra2026agentic,
  author  = {Chhabra, Anshuman and Datta, Shrestha and Nahin, Shahriar Kabir
             and Mohapatra, Prasant},
  title   = {Agentic {AI} Security: Threats, Defenses, Evaluation,
             and Open Challenges},
  journal = {IEEE Access},
  volume  = {14},
  pages   = {49455--49482},
  year    = {2026},
  doi     = {10.1109/ACCESS.2026.3675554}
}

@article{kim2026attack,
  author  = {Kim, Juhee and Liu, Xiaoyuan and Wang, Zhun and Qiu, Shi
             and Li, Bo and Guo, Wenbo and Song, Dawn},
  title   = {The Attack and Defense Landscape of Agentic {AI}:
             A Comprehensive Survey},
  journal = {arXiv preprint arXiv:2603.11088},
  year    = {2026},
  note    = {Accepted to the 35th USENIX Security Symposium (USENIX Security 2026);
             extended version},
  doi     = {10.48550/arXiv.2603.11088}
}

@misc{owasp_agentic_2025,
  author = {{OWASP GenAI Security Project}},
  title  = {{OWASP Top 10 for Agentic Applications}},
  year   = {2025},
  url    = {https://genai.owasp.org/2025/12/09/owasp-top-10-for-agentic-applications-the-benchmark-for-agentic-security-in-the-age-of-autonomous-ai/},
  note   = {Agentic Security Initiative, accessed September 2026}
}

@misc{mitre_atlas,
  author       = {{MITRE}},
  title        = {{ATLAS}: Adversarial Threat Landscape for Artificial-Intelligence Systems},
  howpublished = {\url{https://atlas.mitre.org/}},
  year         = {2026},
  note         = {Version 2026.01, accessed September 2026}
}

@article{bondarenko2025specification,
  author  = {Bondarenko, Alexander and Volk, Denis and Volkov, Dmitrii and Ladish, Jeffrey},
  title   = {Demonstrating Specification Gaming in Reasoning Models},
  journal = {arXiv preprint arXiv:2502.13295},
  year    = {2025},
  doi     = {10.48550/arXiv.2502.13295}
}

@article{cagatan2026reward,
  author  = {{\c{C}}a{\u{g}}atan, {\"O}mer Veysel and Zhao, Xuandong},
  title   = {Reward Hacking in Language Model Agents: Revisiting {AI} Safety Gridworlds},
  journal = {arXiv preprint arXiv:2606.15385},
  year    = {2026},
  doi     = {10.48550/arXiv.2606.15385}
}

@inproceedings{betley2025emergent,
  author    = {Betley, Jan and Tan, Daniel Chee Hian and Warncke, Niels
               and Sztyber-Betley, Anna and Bao, Xuchan and Soto, Mart{\'i}n
               and Labenz, Nathan and Evans, Owain},
  title     = {Emergent Misalignment: Narrow Finetuning Can Produce Broadly Misaligned {LLM}s},
  booktitle = {Proceedings of the 42nd International Conference on Machine Learning},
  series    = {Proceedings of Machine Learning Research},
  volume    = {267},
  pages     = {4043--4068},
  year      = {2025},
  publisher = {PMLR}
}

@article{hubinger2024sleeper,
  author  = {Hubinger, Evan and Denison, Carson and Mu, Jesse and Lambert, Mike
             and Tong, Meg and MacDiarmid, Monte and Lanham, Tamera
             and Ziegler, Daniel M. and Maxwell, Tim and Cheng, Newton
             and Jermyn, Adam and Askell, Amanda and others},
  title   = {Sleeper Agents: Training Deceptive {LLM}s that Persist Through Safety Training},
  journal = {arXiv preprint arXiv:2401.05566},
  year    = {2024},
  doi     = {10.48550/arXiv.2401.05566}
}

@article{greenblatt2024alignment,
  author  = {Greenblatt, Ryan and Denison, Carson and Wright, Benjamin
             and Roger, Fabien and MacDiarmid, Monte and Marks, Sam
             and Treutlein, Johannes and Belonax, Tim and Chen, Jack
             and Duvenaud, David and others},
  title   = {Alignment Faking in Large Language Models},
  journal = {arXiv preprint arXiv:2412.14093},
  year    = {2024},
  doi     = {10.48550/arXiv.2412.14093}
}

@inproceedings{chen2024agentpoison,
  author    = {Chen, Zhaorun and Xiang, Zhen and Xiao, Chaowei and Song, Dawn and Li, Bo},
  title     = {{AgentPoison}: Red-Teaming {LLM} Agents via Poisoning Memory or Knowledge Bases},
  booktitle = {Advances in Neural Information Processing Systems},
  volume    = {37},
  year      = {2024}
}

@article{abuadbba2026human,
  title={Human Society-Inspired Approaches to Agentic AI Security: The 4C Framework},
  author={Abuadbba, Alsharif and Sultan, Nazatul and Nepal, Surya and Jha, Sanjay},
  journal={arXiv preprint arXiv:2602.01942},
  year={2026}
}

@article{srivastava2025memorygraft,
  author  = {Srivastava, Saksham Sahai and He, Haoyu},
  title   = {{MemoryGraft}: Persistent Compromise of {LLM} Agents via Poisoned Experience Retrieval},
  journal = {arXiv preprint arXiv:2512.16962},
  year    = {2025},
  doi     = {10.48550/arXiv.2512.16962}
}

@inproceedings{liu2026memorypropagation,
  author    = {Liu, Hongrui and Xu, Duo and Ma, Qianli and Xu, Shuyang and Qiu, Dong},
  title     = {Memory Poisoning Propagation and Repair Mechanism in Multi-Agent Collaborative Environments},
  booktitle = {Proceedings of the 2nd International Conference on Artificial Intelligence,
               Digital Media Technology and Social Computing},
  pages     = {225--231},
  year      = {2026},
  publisher = {ACM},
  doi       = {10.1145/3806262.3806294}
}

@article{wang2026understanding,
  title={Understanding the Adversarial Landscape of Large Language Models Through the Lens of Attack Objectives},
  author={Wang, Nan and Walter, Kane and Gao, Yansong and Abuadbba, Alsharif},
  journal={IEEE Security \& Privacy},
  volume={24},
  number={1},
  pages={53--60},
  year={2026},
  publisher={IEEE}
}

@article{torra2026memory,
  author  = {Torra, Vicen{\c{c}} and Bras-Amor{\'o}s, Maria},
  title   = {Memory Poisoning and Secure Multi-Agent Systems},
  journal = {arXiv preprint arXiv:2603.20357},
  year    = {2026},
  doi     = {10.48550/arXiv.2603.20357}
}

@article{abuadbba2026promise,
  title={From promise to peril: rethinking cybersecurity red and blue teaming in the age of LLMs},
  author={Abuadbba, Alsharif and Moore, Kristen and Goel, Diksha and Hicks, Chris and Mavroudis, Vasilios and Hasircioglu, Burak and Jennings, Piers},
  journal={IEEE Security \& Privacy},
  volume={24},
  number={2},
  pages={53--63},
  year={2026},
  publisher={IEEE}
}

@inproceedings{zhang2025mcp,
  title={{MCP} Security Bench ({MSB}): Benchmarking Attacks Against Model Context Protocol in {LLM} Agents},
  author={Zhang, Dongsen and Li, Zekun and Luo, Xu and Liu, Xuannan and Li, Peipei and Xu, Wenjun},
  booktitle={The Fourteenth International Conference on Learning Representations},
  year={2026},
  url={https://openreview.net/forum?id=irxxkFMrry}
}

@article{lee2026byzantine,
  author  = {Lee, Haejoon and Yun, Vincent-Daniel and Panagou, Dimitra
             and Karimireddy, Sai Praneeth},
  title   = {Robust Multi-Agent {LLM}s under Byzantine Faults},
  journal = {arXiv preprint arXiv:2605.09076},
  year    = {2026},
  doi     = {10.48550/arXiv.2605.09076}
}

@article{elmir2026byzantine,
  author  = {El Mir, Aya and Tak{\'a}{\v{c}}, Martin and Lahlou, Salem},
  title   = {Byzantine Cheap Talk: Adversarial Resilience and Topology Effects in {LLM} Coordination Games},
  journal = {arXiv preprint arXiv:2606.07790},
  year    = {2026},
  doi     = {10.48550/arXiv.2606.07790}
}

@inproceedings{hu2026lying,
  author    = {Hu, Jinwei and Huang, Xinmiao and Sun, Youcheng and Dong, Yi and Huang, Xiaowei},
  title     = {Lying with Truths: Open-Channel Multi-Agent Collusion for Belief Manipulation via Generative Montage},
  booktitle = {Proceedings of the 64th Annual Meeting of the Association for Computational Linguistics
               (Volume 1: Long Papers)},
  pages     = {5979--5996},
  year      = {2026},
  address   = {San Diego, California, United States},
  publisher = {Association for Computational Linguistics},
  doi       = {10.18653/v1/2026.acl-long.270}
}

@article{zeng2026collusion,
  author  = {Zeng, Xijie and Rudzicz, Frank},
  title   = {Voluntary Collusion with Secret Tools in Competing {LLM} Agents},
  journal = {arXiv preprint arXiv:2605.27593},
  year    = {2026},
  doi     = {10.48550/arXiv.2605.27593}
}

@article{vaswani2017attention,
  title={Attention is all you need},
  author={Vaswani, Ashish and Shazeer, Noam and Parmar, Niki and Uszkoreit, Jakob and Jones, Llion and Gomez, Aidan N and Kaiser, {\L}ukasz and Polosukhin, Illia},
  journal={Advances in neural information processing systems},
  volume={30},
  year={2017}
}

@article{brown2020language,
  title={Language models are few-shot learners},
  author={Brown, Tom and Mann, Benjamin and Ryder, Nick and Subbiah, Melanie and Kaplan, Jared D and Dhariwal, Prafulla and Neelakantan, Arvind and Shyam, Pranav and Sastry, Girish and Askell, Amanda and others},
  journal={Advances in neural information processing systems},
  volume={33},
  pages={1877--1901},
  year={2020}
}

@article{xi2025rise,
  title={The rise and potential of large language model based agents: A survey},
  author={Xi, Zhiheng and Chen, Wenxiang and Guo, Xin and He, Wei and Ding, Yiwen and Hong, Boyang and Zhang, Ming and Wang, Junzhe and Jin, Senjie and Zhou, Enyu and others},
  journal={Science China Information Sciences},
  volume={68},
  number={2},
  pages={121101},
  year={2025},
  publisher={Springer}
}

@inproceedings{zhang2025agent,
  author    = {Zhang, Hanrong and Huang, Jingyuan and Mei, Kai
               and Yao, Yifei and Wang, Zhenting and Zhan, Chenlu
               and Wang, Hongwei and Zhang, Yongfeng},
  title     = {Agent Security Bench ({ASB}): Formalizing and Benchmarking
               Attacks and Defenses in {LLM}-Based Agents},
  booktitle = {International Conference on Learning Representations (ICLR)},
  year      = {2025},
  url       = {https://openreview.net/forum?id=V4y0CpX4hK}
}

@inproceedings{kavathekar2026tamas,
  author    = {Kavathekar, Ishan and Jain, Hemang and Rathod, Ameya
               and Kumaraguru, Ponnurangam and Ganu, Tanuja},
  title     = {{TAMAS}: Benchmarking Adversarial Risks in Multi-Agent
               {LLM} Systems},
  booktitle = {Proceedings of the 64th Annual Meeting of the
               Association for Computational Linguistics
               (Volume 1: Long Papers)},
  pages     = {31238--31268},
  year      = {2026},
  publisher = {Association for Computational Linguistics},
  doi       = {10.18653/v1/2026.acl-long.1442}
}

@article{yin2024safeagentbench,
  author  = {Yin, Sheng and Pang, Xianghe and Ding, Yuanzhuo
             and Chen, Menglan and Bi, Yutong and Xiong, Yichen
             and Huang, Wenhao and Xiang, Zhen and Shao, Jing
             and Chen, Siheng},
  title   = {{SafeAgentBench}: A Benchmark for Safe Task Planning
             of Embodied {LLM} Agents},
  journal = {arXiv preprint arXiv:2412.13178},
  year    = {2024},
  doi     = {10.48550/arXiv.2412.13178}
}

@article{jiang2026agentlab,
  author  = {Jiang, Tanqiu and Wang, Yuhui and Liang, Jiacheng
             and Wang, Ting},
  title   = {{AgentLAB}: Benchmarking {LLM} Agents against
             Long-Horizon Attacks},
  journal = {arXiv preprint arXiv:2602.16901},
  year    = {2026},
  doi     = {10.48550/arXiv.2602.16901}
}

@inproceedings{zou2025security,
  author    = {Zou, Andy and Lin, Maxwell and Jones, Eliot
               and Nowak, Micha and Dziemian, Mateusz and Winter, Nick
               and Nathanael, Valent and Croft, Ayla and Davies, Xander
               and Patel, Jai and Kirk, Robert and Gal, Yarin
               and Hendrycks, Dan and Kolter, J. Zico
               and Fredrikson, Matt},
  title     = {Security Challenges in {AI} Agent Deployment:
               Insights from a Large Scale Public Competition},
  booktitle = {Advances in Neural Information Processing Systems},
  volume    = {38},
  year      = {2025},
  doi       = {10.52202/085713-2676}
}

@article{wang2024survey,
  title={A survey on large language model based autonomous agents},
  author={Wang, Lei and Ma, Chen and Feng, Xueyang and Zhang, Zeyu and Yang, Hao and Zhang, Jingsen and Chen, Zhiyuan and Tang, Jiakai and Chen, Xu and Lin, Yankai and others},
  journal={Frontiers of Computer Science},
  volume={18},
  number={6},
  pages={186345},
  year={2024},
  publisher={Springer}
}

@inproceedings{wu2024autogen,
  title={Autogen: Enabling next-gen LLM applications via multi-agent conversations},
  author={Wu, Qingyun and Bansal, Gagan and Zhang, Jieyu and Wu, Yiran and Li, Beibin and Zhu, Erkang and Jiang, Li and Zhang, Xiaoyun and Zhang, Shaokun and Liu, Jiale and others},
  booktitle={First conference on language modeling},
  year={2024}
}

@article{bandi2025rise,
  title={The rise of agentic ai: A review of definitions, frameworks, architectures, applications, evaluation metrics, and challenges},
  author={Bandi, Ajay and Kongari, Bhavani and Naguru, Roshini and Pasnoor, Sahitya and Vilipala, Sri Vidya},
  journal={Future Internet},
  volume={17},
  number={9},
  pages={404},
  year={2025},
  publisher={MDPI}
}

@article{abou2025agentic,
  title={Agentic AI: a comprehensive survey of architectures, applications, and future directions},
  author={Abou Ali, Mohamad and Dornaika, Fadi and Charafeddine, Jinan},
  journal={Artificial Intelligence Review},
  volume={59},
  number={1},
  pages={11},
  year={2025},
  publisher={Springer}
}

@inproceedings{dong2024survey,
  title={A survey of llm-based agents: Theories, technologies, applications and suggestions},
  author={Dong, Xiaofei and Zhang, Xueqiang and Bu, Weixin and Zhang, Dan and Cao, Feng},
  booktitle={2024 3rd International Conference on Artificial Intelligence, Internet of Things and Cloud Computing Technology (AIoTC)},
  pages={407--413},
  year={2024},
  organization={IEEE}
}

@inproceedings{yao2022react,
  title={React: Synergizing reasoning and acting in language models},
  author={Yao, Shunyu and Zhao, Jeffrey and Yu, Dian and Du, Nan and Shafran, Izhak and Narasimhan, Karthik R and Cao, Yuan},
  booktitle={The eleventh international conference on learning representations},
  year={2022}
}

@article{shinn2023reflexion,
  title={Reflexion: Language agents with verbal reinforcement learning},
  author={Shinn, Noah and Cassano, Federico and Gopinath, Ashwin and Narasimhan, Karthik and Yao, Shunyu},
  journal={Advances in neural information processing systems},
  volume={36},
  pages={8634--8652},
  year={2023}
}

@article{lewis2020retrieval,
  title={Retrieval-augmented generation for knowledge-intensive nlp tasks},
  author={Lewis, Patrick and Perez, Ethan and Piktus, Aleksandra and Petroni, Fabio and Karpukhin, Vladimir and Goyal, Naman and K{\"u}ttler, Heinrich and Lewis, Mike and Yih, Wen-tau and Rockt{\"a}schel, Tim and others},
  journal={Advances in neural information processing systems},
  volume={33},
  pages={9459--9474},
  year={2020}
}

@article{schick2023toolformer,
  title={Toolformer: Language models can teach themselves to use tools},
  author={Schick, Timo and Dwivedi-Yu, Jane and Dess{\`\i}, Roberto and Raileanu, Roberta and Lomeli, Maria and Hambro, Eric and Zettlemoyer, Luke and Cancedda, Nicola and Scialom, Thomas},
  journal={Advances in neural information processing systems},
  volume={36},
  pages={68539--68551},
  year={2023}
}

@misc{AutoGPT,
author = {{Significant Gravitas}},
title = {AutoGPT},
year = {2026},
url = {https://github.com/Significant-Gravitas/AutoGPT},
note = {Accessed: April 17, 2026. Available online: https://github.com/Significant-Gravitas/AutoGPT}
}

@misc{babyagi,
author = {{yoheinakajima}},
title = {BabyAGI},
year = {2026},
url = {https://github.com/yoheinakajima/babyagi},
note = {Accessed: April 17, 2026. Available online: https://github.com/yoheinakajima/babyagi}
}

@inproceedings{liu2024bolaa,
  title={BOLAA: Benchmarking and Orchestrating LLM Autonomous Agents},
  author={Liu, Zhiwei and Yao, Weiran and Zhang, Jianguo and Xue, Le and Heinecke, Shelby and RN, Rithesh and Feng, Yihao and Chen, Zeyuan and Niebles, Juan Carlos and Arpit, Devansh and others},
  booktitle={ICLR 2024 Workshop on Large Language Model (LLM) Agents},
  year={2024}
}

@inproceedings{hong2023metagpt,
  title={MetaGPT: Meta programming for a multi-agent collaborative framework},
  author={Hong, Sirui and Zhuge, Mingchen and Chen, Jonathan and Zheng, Xiawu and Cheng, Yuheng and Wang, Jinlin and Zhang, Ceyao and Wang, Zili and Yau, Steven Ka Shing and Lin, Zijuan and others},
  booktitle={The twelfth international conference on learning representations},
  year={2023}
}

@inproceedings{qian2024chatdev,
  title={Chatdev: Communicative agents for software development},
  author={Qian, Chen and Liu, Wei and Liu, Hongzhang and Chen, Nuo and Dang, Yufan and Li, Jiahao and Yang, Cheng and Chen, Weize and Su, Yusheng and Cong, Xin and others},
  booktitle={Proceedings of the 62nd annual meeting of the association for computational linguistics (volume 1: Long papers)},
  pages={15174--15186},
  year={2024}
}

@article{li2023camel,
  title={Camel: Communicative agents for" mind" exploration of large language model society},
  author={Li, Guohao and Hammoud, Hasan and Itani, Hani and Khizbullin, Dmitrii and Ghanem, Bernard},
  journal={Advances in neural information processing systems},
  volume={36},
  pages={51991--52008},
  year={2023}
}

@inproceedings{du2024improving,
  title={Improving factuality and reasoning in language models through multiagent debate},
  author={Du, Yilun and Li, Shuang and Torralba, Antonio and Tenenbaum, Joshua B and Mordatch, Igor},
  booktitle={Forty-first international conference on machine learning},
  year={2024}
}

@inproceedings{liang2024encouraging,
  title={Encouraging divergent thinking in large language models through multi-agent debate},
  author={Liang, Tian and He, Zhiwei and Jiao, Wenxiang and Wang, Xing and Wang, Yan and Wang, Rui and Yang, Yujiu and Shi, Shuming and Tu, Zhaopeng},
  booktitle={Proceedings of the 2024 conference on empirical methods in natural language processing},
  pages={17889--17904},
  year={2024}
}

@article{liu2023dynamic,
  title={Dynamic llm-agent network: An llm-agent collaboration framework with agent team optimization},
  author={Liu, Zijun and Zhang, Yanzhe and Li, Peng and Liu, Yang and Yang, Diyi},
  journal={arXiv preprint arXiv:2310.02170},
  year={2023}
}

@inproceedings{zhuge2024gptswarm,
  title={Gptswarm: Language agents as optimizable graphs},
  author={Zhuge, Mingchen and Wang, Wenyi and Kirsch, Louis and Faccio, Francesco and Khizbullin, Dmitrii and Schmidhuber, J{\"u}rgen},
  booktitle={Forty-first International Conference on Machine Learning},
  year={2024}
}

@article{narajala2025securing,
  title={Securing agentic ai: A comprehensive threat model and mitigation framework for generative ai agents},
  author={Narajala, Vineeth Sai and Narayan, Om},
  journal={arXiv preprint arXiv:2504.19956},
  year={2025}
}

@article{he2025emerged,
  title={The emerged security and privacy of llm agent: A survey with case studies},
  author={He, Feng and Zhu, Tianqing and Ye, Dayong and Liu, Bo and Zhou, Wanlei and Yu, Philip S},
  journal={ACM Computing Surveys},
  volume={58},
  number={6},
  pages={1--36},
  year={2025},
  publisher={ACM New York, NY}
}

@inproceedings{yu2025survey,
  title={A survey on trustworthy llm agents: Threats and countermeasures},
  author={Yu, Miao and Meng, Fanci and Zhou, Xinyun and Wang, Shilong and Mao, Junyuan and Pan, Linsey and Chen, Tianlong and Wang, Kun and Li, Xinfeng and Zhang, Yongfeng and others},
  booktitle={Proceedings of the 31st ACM SIGKDD Conference on Knowledge Discovery and Data Mining V. 2},
  pages={6216--6226},
  year={2025}
}

@article{guo2025systematic,
  title={Systematic analysis of mcp security},
  author={Guo, Yongjian and Liu, Puzhuo and Ma, Wanlun and Deng, Zehang and Zhu, Xiaogang and Di, Peng and Xiao, Xi and Wen, Sheng},
  journal={arXiv preprint arXiv:2508.12538},
  year={2025}
}

@inproceedings{andriushchenko2024agentharm,
  title     = {{AgentHarm}: A Benchmark for Measuring Harmfulness of {LLM} Agents},
  author    = {Andriushchenko, Maksym and Souly, Alexandra
               and Dziemian, Mateusz and Duenas, Derek
               and Lin, Maxwell and Wang, Justin
               and Hendrycks, Dan and Zou, Andy
               and Kolter, Zico and Fredrikson, Matt
               and Gal, Yarin and Davies, Xander},
  booktitle = {The Thirteenth International Conference on Learning Representations},
  year      = {2025},
  url       = {https://proceedings.iclr.cc/paper_files/paper/2025/hash/c493d23af93118975cdbc32cbe7323f5-Abstract-Conference.html}
}

@article{chiang2025web,
  title={Why are web ai agents more vulnerable than standalone llms? a security analysis},
  author={Chiang, Jeffrey Yang Fan and Lee, Seungjae and Huang, Jia-Bin and Huang, Furong and Chen, Yizheng},
  journal={arXiv preprint arXiv:2502.20383},
  year={2025}
}

@article{hagendorff2026large,
  title={Large reasoning models are autonomous jailbreak agents},
  author={Hagendorff, Thilo and Derner, Erik and Oliver, Nuria},
  journal={Nature Communications},
  year={2026},
  publisher={Nature Publishing Group UK London}
}

@article{mao2026stop,
  title={Stop Fixating on Prompts: Reasoning Hijacking and Constraint Tightening for Red-Teaming LLM Agents},
  author={Mao, Yanxu and Liu, Peipei and Cui, Tiehan and Liu, Congying and Xing, Mingzhe and You, Datao},
  journal={arXiv preprint arXiv:2604.05549},
  year={2026}
}

@article{li2024personal,
  title={Personal llm agents: Insights and survey about the capability, efficiency and security},
  author={Li, Yuanchun and Wen, Hao and Wang, Weijun and Li, Xiangyu and Yuan, Yizhen and Liu, Guohong and Liu, Jiacheng and Xu, Wenxing and Wang, Xiang and Sun, Yi and others},
  journal={arXiv preprint arXiv:2401.05459},
  year={2024}
}

@inproceedings{he2025red,
  title={Red-teaming llm multi-agent systems via communication attacks},
  author={He, Pengfei and Lin, Yuping and Dong, Shen and Xu, Han and Xing, Yue and Liu, Hui},
  booktitle={Findings of the Association for Computational Linguistics: ACL 2025},
  pages={6726--6747},
  year={2025}
}

@inproceedings{greshake2023not,
  title={Not what you've signed up for: Compromising real-world llm-integrated applications with indirect prompt injection},
  author={Greshake, Kai and Abdelnabi, Sahar and Mishra, Shailesh and Endres, Christoph and Holz, Thorsten and Fritz, Mario},
  booktitle={Proceedings of the 16th ACM workshop on artificial intelligence and security},
  pages={79--90},
  year={2023}
}

@article{xiang2024badchain,
  title={Badchain: Backdoor chain-of-thought prompting for large language models},
  author={Xiang, Zhen and Jiang, Fengqing and Xiong, Zidi and Ramasubramanian, Bhaskar and Poovendran, Radha and Li, Bo},
  journal={arXiv preprint arXiv:2401.12242},
  year={2024}
}

@inproceedings{huangresilience,
  title     = {On the Resilience of {LLM}-Based Multi-Agent Collaboration with Faulty Agents},
  author    = {Huang, Jen-Tse and Zhou, Jiaxu and Jin, Tailin
               and Zhou, Xuhui and Chen, Zixi and Wang, Wenxuan
               and Yuan, Youliang and Lyu, Michael and Sap, Maarten},
  booktitle = {Proceedings of the 42nd International Conference on Machine Learning},
  year      = {2025},
  volume    = {267},
  series    = {Proceedings of Machine Learning Research},
  pages     = {26202--26226},
  publisher = {PMLR},
  url       = {https://proceedings.mlr.press/v267/huang25ay.html}
}

@article{gu2024agent,
  title={Agent smith: A single image can jailbreak one million multimodal llm agents exponentially fast},
  author={Gu, Xiangming and Zheng, Xiaosen and Pang, Tianyu and Du, Chao and Liu, Qian and Wang, Ye and Jiang, Jing and Lin, Min},
  journal={arXiv preprint arXiv:2402.08567},
  year={2024}
}

@inproceedings{amayuelas2024multiagent,
  title={Multiagent collaboration attack: Investigating adversarial attacks in large language model collaborations via debate},
  author={Amayuelas, Alfonso and Yang, Xianjun and Antoniades, Antonis and Hua, Wenyue and Pan, Liangming and Wang, William Yang},
  booktitle={Findings of the Association for Computational Linguistics: EMNLP 2024},
  pages={6929--6948},
  year={2024}
}

@article{fang2025we,
  title={We should identify and mitigate third-party safety risks in mcp-powered agent systems},
  author={Fang, Junfeng and Yao, Zijun and Wang, Ruipeng and Ma, Haokai and Wang, Xiang and Chua, Tat-Seng},
  journal={arXiv preprint arXiv:2506.13666},
  year={2025}
}

@article{gasmi2025bridging,
  title={Bridging ai and software security: A comparative vulnerability assessment of llm agent deployment paradigms},
  author={Gasmi, Tarek and Guesmi, Ramzi and Belhadj, Ines and Bennaceur, Jihene},
  journal={arXiv preprint arXiv:2507.06323},
  year={2025}
}

@article{atta2026laaf,
  title={LAAF: Logic-layer Automated Attack Framework A Systematic Red-Teaming Methodology for LPCI Vulnerabilities in Agentic Large Language Model Systems},
  author={Atta, Hammad and Huang, Ken and Lambros, Kyriakos Rock and Mehmood, Yasir and Baig, Zeeshan and Rahman, Mohamed Abdur and Bhatt, Manish and Haq, M and Aatif, Muhammad and Shahzad, Nadeem and others},
  journal={arXiv preprint arXiv:2603.17239},
  year={2026}
}

@article{debenedetti2024agentdojo,
  title={Agentdojo: A dynamic environment to evaluate prompt injection attacks and defenses for llm agents},
  author={Debenedetti, Edoardo and Zhang, Jie and Balunovic, Mislav and Beurer-Kellner, Luca and Fischer, Marc and Tram{\`e}r, Florian},
  journal={Advances in Neural Information Processing Systems},
  volume={37},
  pages={82895--82920},
  year={2024}
}

@inproceedings{zhan2024injecagent,
  title={Injecagent: Benchmarking indirect prompt injections in tool-integrated large language model agents},
  author={Zhan, Qiusi and Liang, Zhixiang and Ying, Zifan and Kang, Daniel},
  booktitle={Findings of the Association for Computational Linguistics: ACL 2024},
  pages={10471--10506},
  year={2024}
}

@article{maloyan2026prompt,
  title={Prompt Injection Attacks on Agentic Coding Assistants: A Systematic Analysis of Vulnerabilities in Skills, Tools, and Protocol Ecosystems},
  author={Maloyan, Narek and Namiot, Dmitry},
  journal={International Journal of Open Information Technologies},
  volume={14},
  number={2},
  pages={1--10},
  year={2026}
}

@article{chang2026overcoming,
  title={Overcoming the Retrieval Barrier: Indirect Prompt Injection in the Wild for LLM Systems},
  author={Chang, Hongyan and Bao, Ergute and Luo, Xinjian and Yu, Ting},
  journal={arXiv preprint arXiv:2601.07072},
  year={2026}
}

@inproceedings{johnson2025dangers,
  title={The Dangers of Indirect Prompt Injection Attacks on LLM-based Autonomous Web Navigation Agents: A Demonstration},
  author={Johnson, Sam and Pham, Viet and Le, Thai},
  booktitle={Proceedings of the 2025 Conference on Empirical Methods in Natural Language Processing: System Demonstrations},
  pages={729--738},
  year={2025}
}

@article{yang2024watch,
  title={Watch out for your agents! investigating backdoor threats to llm-based agents},
  author={Yang, Wenkai and Bi, Xiaohan and Lin, Yankai and Chen, Sishuo and Zhou, Jie and Sun, Xu},
  journal={Advances in Neural Information Processing Systems},
  volume={37},
  pages={100938--100964},
  year={2024}
}

@inproceedings{wang2024badagent,
  title={Badagent: Inserting and activating backdoor attacks in llm agents},
  author={Wang, Yifei and Xue, Dizhan and Zhang, Shengjie and Qian, Shengsheng},
  booktitle={Proceedings of the 62nd Annual Meeting of the Association for Computational Linguistics (Volume 1: Long Papers)},
  pages={9811--9827},
  year={2024}
}

@article{perez2022ignore,
  title={Ignore previous prompt: Attack techniques for language models},
  author={Perez, F{\'a}bio and Ribeiro, Ian},
  journal={arXiv preprint arXiv:2211.09527},
  year={2022}
}

@article{zhang2025udora,
  title={Udora: A unified red teaming framework against llm agents by dynamically hijacking their own reasoning},
  author={Zhang, Jiawei and Yang, Shuang and Li, Bo},
  journal={arXiv preprint arXiv:2503.01908},
  year={2025}
}

@article{zhao2025shadowcot,
  title={Shadowcot: Cognitive hijacking for stealthy reasoning backdoors in llms},
  author={Zhao, Gejian and Wu, Hanzhou and Zhang, Xinpeng and Vasilakos, Athanasios V},
  journal={arXiv preprint arXiv:2504.05605},
  year={2025}
}

@article{kuo2025h,
  title={H-cot: Hijacking the chain-of-thought safety reasoning mechanism to jailbreak large reasoning models, including openai o1/o3, deepseek-r1, and gemini 2.0 flash thinking},
  author={Kuo, Martin and Zhang, Jianyi and Ding, Aolin and Wang, Qinsi and DiValentin, Louis and Bao, Yujia and Wei, Wei and Li, Hai and Chen, Yiran},
  journal={arXiv preprint arXiv:2502.12893},
  year={2025}
}

@article{zhang2025one,
  title={One Token Embedding Is Enough to Deadlock Your Large Reasoning Model},
  author={Zhang, Mohan and Zhang, Yihua and Jia, Jinghan and Wang, Zhangyang and Liu, Sijia and Chen, Tianlong},
  journal={arXiv preprint arXiv:2510.15965},
  year={2025}
}

@article{li2025thinktrap,
  title={ThinkTrap: Denial-of-Service Attacks against Black-box LLM Services via Infinite Thinking},
  author={Li, Yunzhe and Wang, Jianan and Zhu, Hongzi and Lin, James and Chang, Shan and Guo, Minyi},
  journal={arXiv preprint arXiv:2512.07086},
  year={2025}
}

@inproceedings{sternak2025automating,
  title={Automating prompt leakage attacks on large language models using agentic approach},
  author={Sternak, Tvrtko and Runje, Davor and Grano{\v{s}}a, Dorian and Wang, Chi},
  booktitle={2025 MIPRO 48th ICT and Electronics Convention},
  pages={99--104},
  year={2025},
  organization={IEEE}
}

@article{dong2025memory,
  title={Memory Injection Attacks on LLM Agents via Query-Only Interaction},
  author={Dong, Shen and Xu, Shaochen and He, Pengfei and Li, Yige and Tang, Jiliang and Liu, Tianming and Liu, Hui and Xiang, Zhen},
  journal={arXiv preprint arXiv:2503.03704},
  year={2025}
}

@inproceedings{wang2025unveiling,
  title={Unveiling privacy risks in llm agent memory},
  author={Wang, Bo and He, Weiyi and Zeng, Shenglai and Xiang, Zhen and Xing, Yue and Tang, Jiliang and He, Pengfei},
  booktitle={Proceedings of the 63rd Annual Meeting of the Association for Computational Linguistics (Volume 1: Long Papers)},
  pages={25241--25260},
  year={2025}
}

@article{lyu2026adam,
  title={ADAM: A Systematic Data Extraction Attack on Agent Memory via Adaptive Querying},
  author={Lyu, Xingyu and He, Jianfeng and Wang, Ning and Hu, Yidan and Li, Tao and Chen, Danjue and Li, Shixiong and Chen, Yimin},
  journal={arXiv preprint arXiv:2604.09747},
  year={2026}
}

@inproceedings{liu2025topology,
  title     = {Topology Matters: Measuring Memory Leakage in Multi-Agent {LLM}s},
  author    = {Liu, Jinbo and Cao, Defu and Wei, Yifei
               and Su, Tianyao and Liang, Yuan and Dong, Yushun
               and Liu, Yan and Zhao, Yue and Hu, Xiyang},
  booktitle = {Findings of the Association for Computational Linguistics: ACL 2026},
  year      = {2026},
  pages     = {39728--39746},
  publisher = {Association for Computational Linguistics},
  doi       = {10.18653/v1/2026.findings-acl.1980},
  url       = {https://aclanthology.org/2026.findings-acl.1980/}
}

@article{qi2024follow,
  title={Follow my instruction and spill the beans: Scalable data extraction from retrieval-augmented generation systems},
  author={Qi, Zhenting and Zhang, Hanlin and Xing, Eric and Kakade, Sham and Lakkaraju, Himabindu},
  journal={arXiv preprint arXiv:2402.17840},
  year={2024}
}

@article{jiang2024rag,
  title={Rag-thief: Scalable extraction of private data from retrieval-augmented generation applications with agent-based attacks},
  author={Jiang, Changyue and Pan, Xudong and Hong, Geng and Bao, Chenfu and Yang, Min},
  journal={arXiv preprint arXiv:2411.14110},
  volume={4},
  year={2024}
}

@article{liu2025exploit,
  title={Exploit tool invocation prompt for tool behavior hijacking in llm-based agentic system},
  author={Liu, Yu and Xie, Yuchong and Luo, Mingyu and Liu, Zesen and Zhang, Zhixiang and Zhang, Kaikai and Li, Zongjie and Chen, Ping and Wang, Shuai and She, Dongdong},
  journal={arXiv e-prints},
  pages={arXiv--2509},
  year={2025}
}

@inproceedings{wang2025mcptox,
  title     = {{MCPTox}: A Benchmark for Tool Poisoning on Real-World {MCP} Servers},
  author    = {Wang, Zhiqiang and Gao, Yichao and Wang, Yanting
               and Liu, Suyuan and Sun, Haifeng and Cheng, Haoran
               and Shi, Guanquan and Du, Haohua and Li, Xiangyang},
  booktitle = {Proceedings of the AAAI Conference on Artificial Intelligence},
  year      = {2026},
  volume    = {40},
  number    = {42},
  pages     = {35811--35819},
  doi       = {10.1609/aaai.v40i42.40895},
  url       = {https://ojs.aaai.org/index.php/AAAI/article/view/40895}
}

@article{jamshidi2025securing,
  title={Securing the Model Context Protocol: Defending LLMs against tool poisoning and adversarial attacks},
  author={Jamshidi, Saeid and Nafi, Kawser Wazed and Dakhel, Arghavan Moradi and Shahabi, Negar and Khomh, Foutse and Ezzati-Jivan, Naser},
  journal={arXiv preprint arXiv:2512.06556},
  year={2025}
}

@article{nguyen2026five,
  title={Five queries are enough: Query-efficient and surrogate-free membership inference attacks on RAG via entailment},
  author={Nguyen, Nguyen Linh Bao and Ma, Wanlun and Vo, Viet and Abuadbba, Alsharif and Fang, Minghong and Zhang, Jun and Xiang, Yang},
  journal={arXiv preprint arXiv:2605.24312},
  year={2026},
  note    = {Accepted to the 35th USENIX Security Symposium (USENIX Security 2026)}
}

@inproceedings{bhatt2025etdi,
  title={Etdi: Mitigating tool squatting and rug pull attacks in model context protocol (mcp) by using oauth-enhanced tool definitions and policy-based access control},
  author={Bhatt, Manish and Narajala, Vineeth Sai and Habler, Idan},
  booktitle={2025 Cyber Awareness and Research Symposium (CARS)},
  pages={1--6},
  year={2025},
  organization={IEEE}
}

@article{acharya2026formal,
  title={A Formal Security Framework for MCP-Based AI Agents: Threat Taxonomy, Verification Models, and Defense Mechanisms},
  author={Acharya, Nirajan and Gupta, Gaurav Kumar},
  journal={arXiv preprint arXiv:2604.05969},
  year={2026}
}

@article{zhao2025mcp,
  title={When mcp servers attack: Taxonomy, feasibility, and mitigation},
  author={Zhao, Weibo and Liu, Jiahao and Ruan, Bonan and Li, Shaofei and Liang, Zhenkai},
  journal={arXiv preprint arXiv:2509.24272},
  year={2025}
}

@article{dehghantanha2026sok,
  title={SoK: The Attack Surface of Agentic AI--Tools, and Autonomy},
  author={Dehghantanha, Ali and Homayoun, Sajad},
  journal={arXiv preprint arXiv:2603.22928},
  year={2026}
}

@article{raza2025trism,
  title={Trism for agentic ai: A review of trust, risk, and security management in llm-based agentic multi-agent systems},
  author={Raza, Shaina and Sapkota, Ranjan and Karkee, Manoj and Emmanouilidis, Christos},
  journal={arXiv preprint arXiv:2506.04133},
  year={2025}
}

@article{ji2026taming,
  title={Taming Various Privilege Escalation in LLM-Based Agent Systems: A Mandatory Access Control Framework},
  author={Ji, Zimo and Wu, Daoyuan and Jiang, Wenyuan and Ma, Pingchuan and Li, Zongjie and Gao, Yudong and Wang, Shuai and Li, Yingjiu},
  journal={arXiv preprint arXiv:2601.11893},
  year={2026}
}

@article{mitra2026agenticcyops,
  title={AgenticCyOps: Securing Multi-Agentic AI Integration in Enterprise Cyber Operations},
  author={Mitra, Shaswata and Patel, Raj and Mittal, Sudip and Rahman, Md Rayhanur and Rahimi, Shahram},
  journal={arXiv preprint arXiv:2603.09134},
  year={2026}
}

@article{nguyen2025penetration,
  title={Penetration Testing of Agentic AI: A Comparative Security Analysis Across Models and Frameworks},
  author={Nguyen, Viet K and Husain, Mohammad I},
  journal={arXiv preprint arXiv:2512.14860},
  year={2025}
}

@article{lazer2026survey,
  title={A Survey of Agentic AI and Cybersecurity: Challenges, Opportunities and Use-case Prototypes},
  author={Lazer, Sahaya Jestus and Aryal, Kshitiz and Gupta, Maanak and Bertino, Elisa},
  journal={arXiv preprint arXiv:2601.05293},
  year={2026}
}

@article{nakamura2026colosseum,
  title={Colosseum: Auditing Collusion in Cooperative Multi-Agent Systems},
  author={Nakamura, Mason and Kumar, Abhinav and Das, Saswat and Abdelnabi, Sahar and Mahmud, Saaduddin and Fioretto, Ferdinando and Zilberstein, Shlomo and Bagdasarian, Eugene},
  journal={arXiv preprint arXiv:2602.15198},
  year={2026}
}

@article{sivaroopan2026shield,
  title={SHIELD: An Auto-Healing Agentic Defense Framework for LLM Resource Exhaustion Attacks},
  author={Sivaroopan, Nirhoshan and Thilakarathna, Kanchana and Zomaya, Albert and Guo, Yi and Plested, Jo and Lynar, Tim and Yang, Jack and Yang, Wangli and others},
  journal={arXiv preprint arXiv:2601.19174},
  year={2026}
}

@article{anbiaee2026security,
  title={Security Threat Modeling for Emerging AI-Agent Protocols: A Comparative Analysis of MCP, A2A, Agora, and ANP},
  author={Anbiaee, Zeynab and Rabbani, Mahdi and Mirani, Mansur and Piya, Gunjan and Opushnyev, Igor and Ghorbani, Ali and Dadkhah, Sajjad},
  journal={arXiv preprint arXiv:2602.11327},
  year={2026}
}

@article{qu2026supply,
  title={Supply-Chain Poisoning Attacks Against LLM Coding Agent Skill Ecosystems},
  author={Qu, Yubin and Liu, Yi and Geng, Tongcheng and Deng, Gelei and Li, Yuekang and Zhang, Leo Yu and Zhang, Ying and Ma, Lei},
  journal={arXiv preprint arXiv:2604.03081},
  year={2026}
}

@article{jiang2026agentic,
  title={Agentic AI as a Cybersecurity Attack Surface: Threats, Exploits, and Defenses in Runtime Supply Chains},
  author={Jiang, Xiaochong and Yang, Shiqi and Yang, Wenting and Liu, Yichen and Ji, Cheng},
  journal={arXiv preprint arXiv:2602.19555},
  year={2026}
}

@article{abdennebi2026lang,
  title={LanG--A Governance-Aware Agentic AI Platform for Unified Security Operations},
  author={Abdennebi, Anes and Kara, Nadjia and Lahlou, Laaziz and Ould-Slimane, Hakima},
  journal={arXiv preprint arXiv:2604.05440},
  year={2026}
}

@article{singh2026shifting,
  title={Shifting from Injection to Interaction: Rethinking Web Security in the Age of LLMs and Beyond},
  author={Singh, Nivedita and Abuadbba, Alsharif and Gao, Yansong and Nepal, Surya and Kim, Hyoungshick},
  journal={arXiv preprint arXiv:2609.03999},
  year={2026}
}
